\documentclass[aps,prl,reprint,superscriptaddress,nofootinbib]{revtex4-2}
\usepackage{amsmath,amsfonts,amssymb,bm}
\usepackage{graphicx}
\usepackage{bbold}
\usepackage[normalem]{ulem}
\usepackage[dvipsnames,usenames]{xcolor}
\usepackage[colorlinks,citecolor=blue]{hyperref}
\usepackage{orcidlink}

\newcounter{defcountereq}
\newcounter{defcounterfig}
\newcounter{defcountertab}
\newenvironment{sequation}{
\addtocounter{equation}{-1}
\refstepcounter{defcountereq}

\begin{equation}}{\end{equation}}

\newenvironment{sfigure}{
\addtocounter{figure}{-1}
\refstepcounter{defcounterfig}

\begin{figure}}{\end{figure}}

\newenvironment{sfigure*}{
\addtocounter{figure}{-1}
\refstepcounter{defcounterfig}

\begin{figure*}}{\end{figure*}}

\begin{document}

\title{Robust Integration of Fast Flavor Conversions in Classical Neutrino Transport}

\author{Zewei Xiong\orcidlink{0000-0002-2385-6771}}
\email{z.xiong@gsi.de}
\affiliation{GSI Helmholtzzentrum {f\"ur} Schwerionenforschung, Planckstra{\ss}e 1, 64291 Darmstadt, Germany}

\author{Meng-Ru Wu\orcidlink{0000-0003-4960-8706}}
\email{mwu@gate.sinica.edu.tw}
\affiliation{Institute of Physics, Academia Sinica, Taipei 115201, Taiwan}
\affiliation{Institute of Astronomy and Astrophysics, Academia Sinica, Taipei 106319, Taiwan}
\affiliation{Physics Division, National Center for Theoretical Sciences, Taipei 106319, Taiwan}

\author{Manu George\orcidlink{0000-0002-8974-5986}}
\email{manug@gate.sinica.edu.tw}
\affiliation{Institute of Physics, Academia Sinica, Taipei 115201, Taiwan}

\author{Chun-Yu Lin\orcidlink{0000-0002-7489-7418}}
\email{lincy@nchc.org.tw}
\affiliation{National Center for High-performance Computing, Hsinchu 30076, Taiwan}

\date{\today}

\begin{abstract}
The quantum kinetic evolution of neutrinos in dense environments, such as the core-collapse supernovae or the neutron star mergers, can result in fast flavor conversion (FFC), presenting a significant challenge to achieving robust astrophysical modeling of these systems. 
Recent works that directly simulate the quantum kinetic transport of neutrinos in localized domains have suggested that the asymptotic outcome of FFCs can be modeled by simple analytical prescriptions when coarse grained over a size much larger than the FFC length scale. 
In this \emph{Letter}, by leveraging such a scale separation, we incorporate the analytical prescriptions into global simulations that solve the classical neutrino transport equation including collisions and advection under spherical symmetry. 
We demonstrate that taking this approach allows to obtain results 
that \emph{quantitatively} agree with those directly from the corresponding global quantum kinetic simulations
and precisely capture the collisional feedback effect for cases where the FFC happens inside the neutrinosphere. 
Notably, the effective scheme does not require resolving the FFC time and length scales, hence only adds negligible computational overhead to classical transport.  
Our work highlights that efficient and robust integration of FFCs in classical neutrino transport used in astrophysical simulation can be feasible.
\end{abstract}

\maketitle
% \end{CJK*}
\graphicspath{{./}{figures/}}

{\it Introduction---}
The robustness of astrophysical simulations for explosive events such as core-collapse supernovae (CCSNe) and neutron star mergers (NSMs) relies on the accurate numerical modeling of the nonequilibrium classical neutrino transport from the neutrino-trapped to free-streaming regime~\cite{janka2012explosion,Pan2018theimpact,Burrows2016crucial,Richers2017detailed,Nagakura:2017mnp,oconnor2018global,Mezzacappa2020physical,Foucart2022neutrino}.  
However, recent studies have revealed that the quantum mechanical nature of neutrinos, as manifested in their flavor oscillations, can lead to various collective modes of flavor conversions in regions where neutrinos decouple from matter~\cite{capozzi2022neutrino,volpe2023neutrinos,Fischer:2023ebq}. 
These collective phenomena, originated from the nonlinear self-interaction of neutrino forward scattering, may significantly affect the theoretical modeling and the aftermath of CCSNe and NSMs, thus emphasizing the need for quantum kinetic transport of neutrinos~\cite{sigl1993general,vlasenko2014neutrino,volpe2015neutrino,Kartavtsev2015neutrino,blaschke2016neutrino,Richers2019neutrino,fiorillo2024inhomogenous}. 

In particular, the so-called fast flavor instability (FFI)~\cite{sawyer2009multiangle,izaguirre2017fast,capozzi2017fast,yi2019dispersion} can exist at locations where the angular distribution of the neutrino electron lepton number (ELN) density contains a crossing, i.e., the ELN angular distribution function changes sign~\cite{morinaga2022fast,dasgupta2022collective}, and is commonly found in the interior of CCSNe and NSM remnants~\cite{wu2017fast,wu2017imprints,azari2019linear,azari2020fast,nagakura2019fast,morinaga2020fast,abbar2020fast,george2020fast,glas2020fast,nagakura2021occurrence,abbar2021characteristics,harada2022prospects,richers2022evaluating}.
The presence of FFI can lead to the emergence of fast flavor conversion (FFC) and alter the neutrino flavor content within a timescale of subnanosecond and a length scale of subcentimeter, much shorter than the typical hydrodynamical or interaction time and length scales governed by classical processes. 

The large separation of scales implies an exceedingly high computational cost for a direct implementation of neutrino quantum kinetic transport in simulations of CCSNe and NSMs. 
Intensive studies aimed to obtain solutions to this problem have been explored. 
A large number of recent works have numerically solved the neutrino quantum kinetic equation ($\nu$QKE) in localized simulation domains to understand the outcome of FFCs and found that the ELN crossings are erased on a coarse-grained sense~\cite{martin2020dynamic,bhattacharyya2021fast,bhattacharyya2020late,wu2021collective,richers2021particle,richers2021neutrino,zaizen2021nonlinear,abbar2022suppression,richers2022code,bhattacharyya2022elaborating,grohs2023neutrino,zaizen2023simple,zaizen2023characterizing,xiong2023evaluating,nagakura2023bgk,grohs2024two,froustey2024neutrino}. 
Attempts to model the coarse-grained FFC outcome based on these simulations using analytical formulas~\cite{xiong2021stationary,bhattacharyya2021fast,zaizen2023simple,xiong2023evaluating} or through the use of machine learning~\cite{abbar2024physics,abbar2024applications} have proven that accurate  
outcomes relevant to the modeling of neutrino transport can be obtained. 
Global transport simulations solving $\nu$QKE with different simplifications, e.g., using artificially quenched neutrino self-interaction or adopting spatial resolutions lower than the FFC length scales, have been performed~\cite{capozzi2019collisional,nagakura2022time,xiong2023evolution,nagakura2023basic,nagakura2023roles,nagakura2023global,xiong2024fast,shalgar2023supernova,shalgar2023neutrino,cornelius2024perturbing} and provided improved understanding of the emergence and evolution of FFCs in static background. 
On the other hand, preliminary efforts to include the FFC outcome in parametric fashions in hydrodynamical simulations equipped with classical neutrino transport solvers were conducted in \cite{li2021neutrino,just2022fast,fernandez2023fast,ehring2023fast,ehring2023fast2}.
Meanwhile, new theoretical frameworks including the subgrid methods, ``miscidynamics,'' and quasilinear approach have been proposed in~\cite{johns2023thermodynamics,nagakura2023bgk,johns2024subgrid,fiorillo2024fast,fiorillo2024theory}. 

Despite all these tremendous efforts, an important question that remains to be answered is whether or not the coarse-grained FFC outcome learned from local $\nu$QKE simulations can be effectively included in classical transport used in astrophysical simulations based on the principle of separation of scales. 
In this Letter, we tackle this question by directly incorporating the coarse-grained FFC prescriptions into global simulations that solve the classical neutrino transport equation under spherical symmetry. 
For the first time, we show \emph{robust} agreement between results from this approach and those directly solving the $\nu$QKE in our earlier work~\cite{xiong2024fast}, which reveals the viability of effective integration of FFC in astrophysical simulations.

{\it Model---}
We use the extended version~\cite{xiong2024fast} of \textsc{cose$\nu$}~\cite{george2023cosenu} to numerically solve the $\nu$QKE and the corresponding effective classical transport (ECT) under static and spherically symmetric background using discrete ordinates method in Minkowski spacetime.
The code includes the effective Hamiltonian terms governing flavor oscillations of neutrinos and their interactions with matter and can reproduce the general properties of classical neutrino transport results obtained from SN simulations using the \textsc{agile-boltztran} code~\cite{mezzacappa1993numerical,liebendorfer2001conservative,liebendorfer2004finite}.
All $\nu$QKE and the ECT simulations are performed in a radial range from $r=20$~km to 80~km, spanning both the neutrino-trapped and free-streaming regimes for spherically symmetric CCSN background profiles based on a 25$M_\odot$ progenitor as used in \cite{xiong2024fast} with two-flavor scheme. 
We focus on Models II, III, and IV of \cite{xiong2024fast}\footnote{The neutrino angular distributions in Model I of \cite{xiong2024fast} does not contain any ELN crossings, i.e., no FFC, and will not be further discussed below.}. 
These models take different electron fraction profiles, which result in different number density ratios of $n_{\bar\nu_e}$/$n_{\nu_e}$ and ELN angular distributions. 
In Models II and III, $n_{\nu_e}\gtrsim n_{\bar\nu_e}$ at all radii without 
FFCs, while $n_{\nu_e}\lesssim n_{\bar\nu_e}$ in Model IV, representing the $\nu_e$ and $\bar\nu_e$ dominating cases, respectively (see \hyperref[SM]{Supplemental Material}).
Moreover, the FFIs exist inside the neutrinospheres in Models III and IV, which leads to an intriguing collisional feedback effect that continuously alters the evolution of neutrino distributions after the prompt FFC.  

To perform $\nu$QKE simulations within a reasonable time frame, we take an attenuation function $a_{\nu\nu}(r)$ to quench the strength of the neutrino-neutrino forwarding scattering Hamiltonian, $\mathbf{H_{\nu\nu}}(v_r)= \sqrt{2} G_F \int d E'\, d v_r' (1-v_r v_r') [\varrho(E',v_r')-\bar\varrho^*(E',v_r')]$ with $G_F$ the Fermi constant, $E$ and $v_r$ the energy and the radial velocity of neutrinos, and $\varrho$ and $\bar\varrho$ the density matrices of neutrino and antineutrinos.
Natural units with  $\hbar=c=1$ are adopted throughout this Letter. 
The diagonal elements of $\varrho$ and $\bar\varrho$ are related to
$n_{\nu_i}(r, t) = \int d E\, d v_r\, \varrho_{ii}$ and $n_{\bar\nu_i}(r, t) = \int d E\, d v_r\, \bar\varrho_{ii}$ for $i=e,\,x$ with $x$ denoting the heavy-lepton flavor.
The attenuation function takes the form $a_{\nu\nu}(r) = a/[1+e^{(30-r\,[{\rm km}])/2.5}]$, such that $a_{\nu\nu}(r)\approx a$  at $r>35$~km and $a_{\nu\nu}(r)> 0.1a$ at $r>25$~km for the radial range where FFC occurs. 
We take two different attenuation factors $a=4\times 10^{-3}$ and $a=10^{-3}$. 
Even with the quenched $\mathbf{H_{\nu\nu}}$, the $\nu$QKE simulations need 50,000 (25,000) uniform radial grids for larger (smaller) value of $a$, labeled as the ``$\nu$QKE-H'' (``$\nu$QKE-L'') models in the rest of the Letter to resolve the attenuated FFC scale. 
For the time step size, we take $\Delta t=1.6$~ns ($3.2$~ns) for the ``$\nu$QKE-H'' (``$\nu$QKE-L'') models. 
Details of simulation parameters are given in \hyperref[SM]{Supplemental Material}. 

For the ECT simulations, 
we assume that when the FFI exists at any radius, i.e., when there exists an ELN angular crossing, the FFC quickly relaxes the local neutrino flavor content to a quasistationary state when averaging over a length scale much larger than the FFC one, within a timescale much shorter than the collisional and the advection timescales. 
In the quasistationary state, the coarse-grained ELN angular distribution does not contain any crossings~\cite{bhattacharyya2021fast,wu2021collective,zaizen2023simple,xiong2023evaluating,fiorillo2024fast}, 
and can be approximated by certain analytical prescriptions, where flavor equilibration takes place in a part of the angular domain~\cite{xiong2023evaluating,zaizen2023simple}, with the off-diagonal elements of $\varrho$ and $\bar\varrho$ averaging to $\sim 0$.  
Based on the above assumptions rooted from the scale separation nature of the problem, we then take the following practical procedures to effectively incorporate the FFC in the classical transport. 

\begin{figure*}[!hbt]
\includegraphics[width=\textwidth]{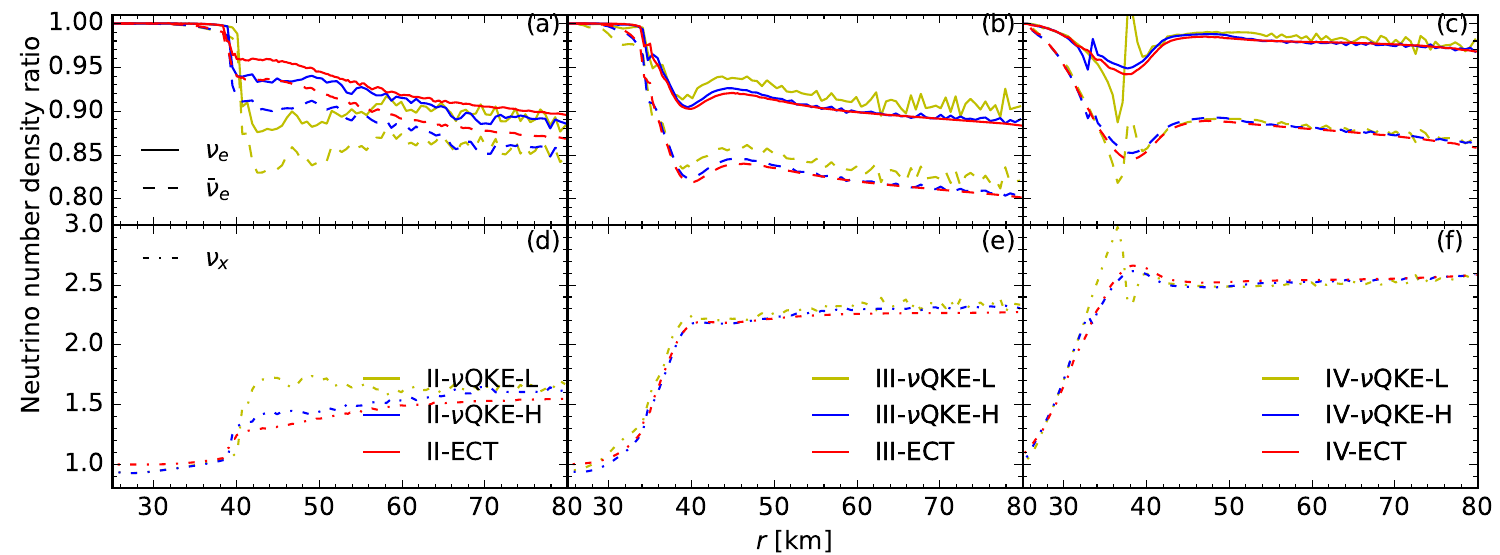}
\caption{\label{fig:rho_ratio_r} Comparison of the radial profiles of neutrino number density ratios $n_\nu^{\rm FFC}/n_\nu^{\rm NFC}$ obtained in $\nu$QKE simulations using different attenuation factors $a=10^{-3}$ (yellow, labeled by ``$\nu$QKE-L'') and $a=4\times 10^{-3}$ (blue, labeled by ``$\nu$QKE-H'') with the corresponding classical transport simulation outcome taking into account the FFC effectively (red, labeled by ``ECT'') for $\nu_e$, $\bar\nu_e$ [panels (a)--(c)] and $\nu_x$ [panels (d)--(f)] in Models II--IV, respectively.
}
\end{figure*}

First, the coarse-grained nature allows (and demands) us to take a radial grid size much larger than 
the size of $\mathbf{H_{\nu\nu}}$. 
Thus, in all our ECT 
models, we use 250 radial grids in the same simulation domain and a larger
$\Delta t=32$~ns. 
Second, we only evolve the diagonal elements $\varrho_{ii}$ and $\bar\varrho_{ii}$ by setting the coherent propagation Hamiltonian term, as well as the off-diagonal entries in the collision term to zero. 
Third, at each time step, we check whether there exist angular crossings in the ELN function $G(v_r)= \langle \varrho_{ee} \rangle_E-\langle \bar\varrho_{ee} \rangle_E-\langle \varrho_{xx} \rangle_E+\langle \bar\varrho_{xx} \rangle_E$, where $\langle\cdot\rangle_E$ denotes the energy-integrated quantities at all radial grids.
If an ELN crossing is found at a given radius, we then assume that immediate flavor redistribution takes place due to FFC and apply the power-1/2 prescription scheme derived in \cite{xiong2023evaluating} to calculate the redistributed diagonal elements in $\varrho$ and $\bar\varrho$. 
Specifically, under the two-flavor scenario, the energy-independent survival probability for (anti)neutrinos is given by
\begin{equation}\label{eq:p_sur}
    P(v_r,w) = 
    \begin{cases}
        \frac{1}{2} & {\rm for~} v_<, \\
        1-\frac{1}{2}h(|v_r-v_c|/w) & {\rm for~} v_>,
    \end{cases}
\end{equation}
where the function $h(x)=(x^2+1)^{-1/2}$ and $v_c$ is the crossing velocity at which $G(v_c)=0$. 
In Eq.~\eqref{eq:p_sur}, $v_<$ ($v_>$) is defined as the $v_r$ range
over which the absolute value of the positive or negative ELN, $|I_+|$ or $|I_-|$, is smaller (larger).
The quantities $I_\pm$ are defined as $\int dv_r G(v_r) \Theta[\pm G(v_r)]$, with $\Theta$ denoting the Heaviside function. 
The parameter $w$ can be calculated so that the conservation of ELN holds, i.e., $\int dv_r [2P(v_r,w)-1] G(v_r)=I_+ + I_-$.
Finally, with Eq.~\eqref{eq:p_sur}, the neutrino flavor content $\varrho^{\rm (f)}_{ii}$ after the FFC are replaced by 
\begin{align}
\varrho_{ee} & \rightarrow \varrho_{ee}^{\rm (f)}=\varrho_{ee}P+\varrho_{xx}(1-P), \nonumber \\
\varrho_{xx} & \rightarrow \varrho_{xx}^{\rm (f)}=\varrho_{ee}(1-P)+\varrho_{xx}P, 
\end{align}
for all energy grids at the radial grids where sizable ELN crossings with $(I_</I_>) >10^{-3}$ are found, where $I_<={\rm min}(|I_+|,|I_-|)$ and $I_>={\rm max}(|I_+|,|I_-|)$.
A similar redistribution applies to antineutrinos as well. 
We also note that Eq.~\eqref{eq:p_sur} only applies to cases with at most a single angular crossing at any given radius, which is true in all models examined in this work. 
Besides the procedures outlined above, all other settings in the 
ECT simulations are identical to those in the corresponding $\nu$QKE simulations. 

We run all the $\nu$QKE and the ECT simulations 
up to $320~\mu$s when the systems have settled into the asymptotic states. 
The large reduction of radial grid numbers and the increased $\Delta t$ in the ECT simulations significantly reduce the computation time by $\gtrsim\mathcal{O}(10^3)$ from the $\nu$QKE runs for all cases (see \hyperref[SM]{Supplemental Material}). 
The $\nu$QKE results shown in all figures are coarse-grain averaged over a size of 0.6~km.

\begin{figure}[t]
\includegraphics[width=\columnwidth]{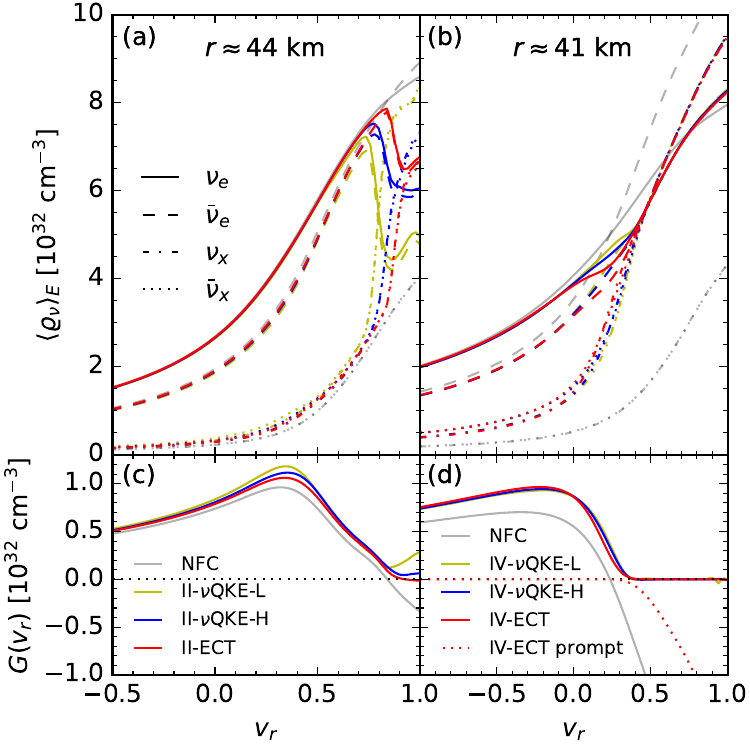}
\caption{\label{fig:an_conv}The neutrino angular distributions for $\nu_e$, $\nu_x$ and their antineutrinos at $r\simeq 44$~km in Model II [panel (a)] and $r\simeq 41$~km in Model IV [panel (b)]. 
The corresponding ELN angular distributions are shown in panels (c) and (d). 
The yellow and blue curves are obtained by the $\nu$QKE simulations and the red curves are from the corresponding ECT models. 
We only plot $v_r$ from -0.5 to 1 in (a) and (c) for better illustration. 
The overlapping asymptotic curves of $\bar\nu_e$ and $\bar\nu_x$ are on the top of those of $\nu_e$ and $\nu_x$ for $v_r>0.4$ in (b). The ELN distribution after the prompt FFC is shown by the dotted red curve in (d).}
\end{figure}

{\it Results---}
Figure~\ref{fig:rho_ratio_r} compares the asymptotic radial profiles of the number density ratio between cases with FFCs and those with no flavor conversion, denoted as $n_\nu^{\rm FFC}/n_\nu^{\rm NFC}$ for all models.
The FFC leads to more enhanced $n_{\nu_x}$ in Model IV, followed by Model III and II in the $\nu$QKE simulations. 
For $\nu_e$ and $\bar\nu_e$, since the FFC takes place at radii inside neutrinospheres in Models III and IV, 
collisions inside neutrinospheres can repopulate their number densities after the prompt phase of FFC and work along with the subsequent FFC to largely alter the ELN distribution 
[see Fig.~\ref{fig:an_conv}(d) and the discussion below]~\cite{xiong2024fast}. 
Consequently, the amount of reduction in $n_{\nu_e}$ and $n_{\bar\nu_e}$ are less so that their number density ratios in Models III and IV become comparable to those in Model II.
All changes of neutrino flavor content can alter their net heating rates and 
affect the fate of SN explosions~\cite{ehring2023fast,ehring2023fast2,nagakura2023roles,xiong2024fast}. 

Comparing the ECT results to the $\nu$QKE ones, it clearly shows remarkably \emph{quantitative} agreement in all three models.  
In particular, the differences in number density ratios between the ECT and $\nu$QKE-H models are only up to $\lesssim 5\%$ in Model II, and are even smaller than $2\%$ for most radii in Models III and IV, where the collisional feedback effects are important.
For the $\nu$QKE-L models, slightly larger differences exist around $r\simeq 45$~km in Model II and $r\simeq 37$~km in Model IV. 

The differences in Model II are primarily associated with the flavor overconversion of the forward propagating neutrinos during the FFC, which occurs before the ELN distribution settles down to the quasistationary state that can be approximated by Eq.~\eqref{eq:p_sur}. 
With the lower attenuation factor applied in the $\nu$QKE-L model (more attenuated), the lengthened FFC timescale prevents FFC from reaching the quasistationary state before advection takes effect.
Consequently, a larger amount of flavor overconversion is imprinted in the asymptotic state in Model II-$\nu$QKE-L than in Model II-$\nu$QKE-H.
In Model IV, the 
negative ELN dominated by $\bar\nu_e$'s before FFCs changes to positive due to the collisional feedback.
The strong attenuation of $\mathbf{H}_{\nu\nu}$ in \mbox{$\nu$QKE-L} model can cause the ELN to switch back to negative at $r\sim 30$--37~km, resulting in a sharp transition at $r\simeq 37$~km.
However, in the $\nu$QKE-H model where less attenuation is applied, the transition becomes less sharp and the location is shifted inward to $r\simeq 33$~km.
This indicates that the difference is related to the artificial attenuation and the modified ratios between the collisional 
and FFC rates.
Overall, the better agreement between the ECT and $\nu$QKE-H models than in between the ECT and $\nu$QKE-L models suggests that in the $\nu$QKE-L models the larger attenuation of $\mathbf{H}_{\nu\nu}$ may have quenched the FFC rate too much, particularly in Models II-$\nu$QKE-L and IV-$\nu$QKE-L.

To further illustrate the increasing agreement between the 
ECT and the $\nu$QKE results with less attenuation, we 
compare the asymptotic neutrino angular distributions obtained at $r\simeq 44$~km in Model II in Fig.~\ref{fig:an_conv}(a) and $r\simeq 41$~km in Model IV in Fig.~\ref{fig:an_conv}(b). 
In Model II-ECT, the reinforced flavor redistribution scheme leads to nearly equal amounts of electron and heavy lepton flavors in both neutrino and antineutrino sectors in the velocity range $v_r\gtrsim v_c$. 
In both Models II-$\nu$QKE-L and II-$\nu$QKE-H, slight flavor overconversion is obtained.
In Model IV, although FFCs constrain the ELN distribution by prohibiting the reappearance of new crossing, collisions can still affect the angular distribution of each neutrino species and exert feedback.
This leads to the gradual change of ELN distribution and $v_c$, so that flavor equilibration is reached at $v_r>v_c$ in the asymptotic state, in contrast to the range $v_r<v_c$ indicated by the red dotted curve in Fig.~\ref{fig:an_conv}(d).
Such an effect can be well reproduced in ECT models as illustrated by Figs.~\ref{fig:an_conv}(b) and \ref{fig:an_conv}(d).
Once again, the results in the $\nu$QKE-H models clearly agree with those in the ECT models better than the $\nu$QKE-L models.

\begin{figure}[t]
\includegraphics[width=\columnwidth]{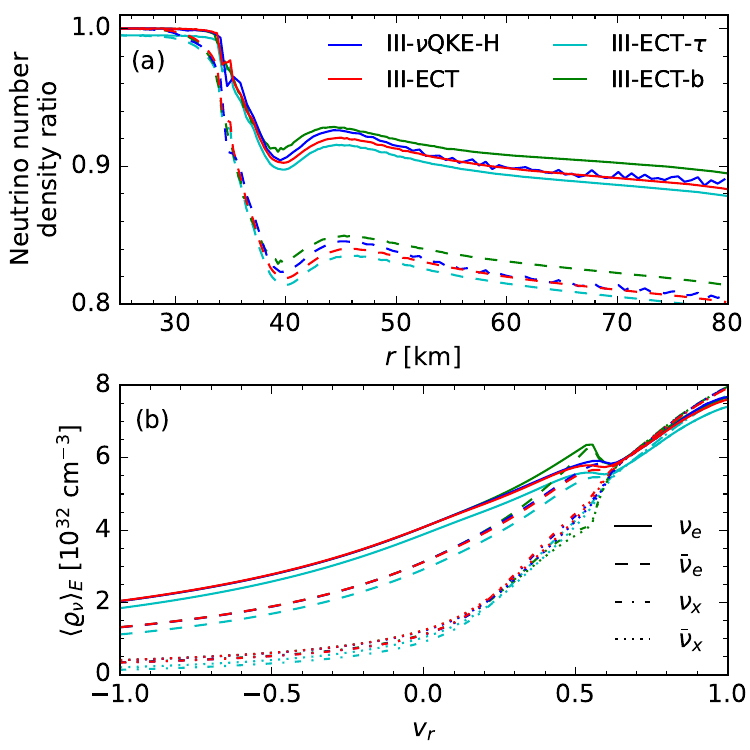}
\caption{\label{fig:box} 
The radial profiles of neutrino number density ratios $n_\nu^{\rm FFC}/n_\nu^{\rm NFC}$ 
[panel (a)] and the angular distributions at $r\simeq 41$~km [panel (b)] in Model III-$\nu$QKE-H (blue), III-ECT (red), \mbox{III-ECT}-$\tau$ (cyan), and III-ECT-b (green). 
The ECT-$\tau$ model takes into account the FFC timescale.
The ECT-b model uses a different flavor redistribution scheme from the ECT model.
As the curves for Model III-ECT-$\tau$ completely overlap with those from the ECT model, they are shifted by -0.005 in (a) and -0.2 in (b), respectively.
}
\end{figure}

{\it Alternative effective schemes---}
Besides the fiducial effective model used above, we have also performed two additional ECT schemes. 
The first set (labeled by ``ECT-$\tau$'') is to involve an FFC timescale to relax the assumption of instantaneous flavor redistribution~\cite{nagakura2023bgk}. 
The characteristic FFC timescale may be estimated by $\tau=G_F^{-1}|I_+ I_-|^{-1/2}$, which is proportional to the geometrical mean of the positive and negative ELNs~\cite{nagakura2019fast}.  
At each time step, we compute $\tau$ at where the ELN crossings are found and update the neutrino and antineutrino flavor content by $\varrho+(\varrho^{\rm (f)}-\varrho)\,{\rm min}(\Delta t/\tau, 1)$ and $\bar\varrho+(\bar\varrho^{\rm (f)}-\bar\varrho)\,{\rm min}(\Delta t/\tau, 1)$, respectively. 
When $\tau<\Delta t$, it reduces back to the instantaneous redistribution scheme.
The results obtained with this method are nearly identical to those with instantaneous flavor redistribution in all models; see e.g., Fig.~\ref{fig:box} for Model III.
This comparis%FFCon supports the validity of implementing instantaneous flavor redistribution in hydrodynamic simulations~\cite{li2021neutrino,just2022fast,fernandez2023fast,ehring2023fast,ehring2023fast2}.

The second set of additional simulations is to use the ``box'' prescription \cite{zaizen2023simple} by setting $P(v_r)=1-I_</(2I_>)$ for $v_>$, labeled by ``ECT-b,'' which offers easier implementation but creates discontinuities in 
neutrino angular distributions~\cite{xiong2023evaluating}. 
Interestingly, despite the presence of local discontinuities [see, e.g., Fig.~\ref{fig:box}(b)], the resulting neutrino densities do not differ substantially from those using the power-1/2 prescription, as shown in Fig.~\ref{fig:box}(a) for Model III. 
We have also examined that for other prescriptions, which model the FFC outcome by continuous functions in~\cite{xiong2023evaluating}, the results are nearly as good as those with the power-1/2 scheme.

{\it Summary and discussions---}
We have performed the first comparison of the global $\nu$QKE simulations to the corresponding 
ECT simulations that effectively take into account the outcome of the FFC.
Utilizing the scale separation between the FFC and the neutrino collisions as well as advection, we have implemented the analytical prescriptions proposed in \cite{xiong2023evaluating} as the instantaneous FFC outcome over a radial size much larger than the FFC length scale into global simulation of classical neutrino transport that includes collisions and advection.  
The ECT scheme takes significantly reduced grid number and time step than required by the $\nu$QKE simulations to obtain results that quantitatively agree with those by solving the full $\nu$QKE with differences of less than a few percents.  
The models that we examined in this work cover different cases with number densities dominated by $\nu_e$ or $\bar\nu_e$, and scenarios where the FFIs 
exist inside and outside neutrinospheres.
In particular, the evolution of neutrino flavor content driven by the collisional feedback when the FFIs are located inside the neutrinospheres 
can also be robustly modeled by the 
ECT method.

Moreover, we have found increasing agreement between the $\nu$QKE results with less attenuation in the neutrino-neutrino forward scattering Hamiltonian with the results obtained by the 
ECT method. 
Taking into account the characteristic FFC timescale or using different analytical prescriptions to describe the redistribution of neutrinos by FFC also gives rise to similar results in the ECT models. 
Our results suggest that the flavor redistribution obtained on a coarse-grained level achieved in the FFC timescale in local FFC simulations with periodic boundary conditions can be used as a good proxy in global transport simulations that aim to include the impact of FFCs. 
Although the models examined are based on a particular SN simulation profile, we fully expect that the method is generally applicable to different SN backgrounds so long as the separation of scales holds at where FFC occurs. 
Also, due to the local nature of the ECT scheme, its implementation does not add significant computational burden to classical transport. 
Thus, this work serves as a proof-of-principle study for the upcoming efforts to accurately identify the FFIs and integrate the FFC outcome utilizing machine learning and neural networks into the CCSN and NSM simulations equipped with classical two-moment neutrino transport schemes. 

Despite the remarkable success presented above, it remains desired to explore whether flavor swap through the electron lepton zero surface~\cite{nagakura2023global,zaizen2023fast} or 
near-quasistationary flavor transformation due to the temporal evolution of the environment~\cite{johns2023thermodynamics,johns2024subgrid,fiorillo2024fast} can be treated effectively as well. 
If a non-negligible amount of flavor overconversion due to FFC may exist for certain ELN distributions that cannot be described by analytical prescriptions, it remains to be seen whether machine-learning based methods can accurately predict the related FFC outcome. 
All these as well as the generalization to multidimensional simulations and further development of ECT models for other types of collective oscillations are beyond the scope of this work but will be pursued in upcoming studies.

\begin{acknowledgments}
{\it Acknowledgments---}We thank Tobias Fischer and Noshad Khosravi Largani for providing the supernova profile. 
We are grateful for comments from
Sajad Abbar, Damiano Fiorillo, Luke Johns, Gabriel Mart\'{i}nez-Pinedo, Hiroki Nagakura, Georg Raffelt, and Irene Tamborra.
Z.X. acknowledges support of the European Research Council (ERC) under the European Union’s Horizon 2020 research and innovation program (ERC Advanced Grant KILONOVA No. 885281) and under the ERC Starting Grant (NeuTrAE, No. 101165138), the Deutsche Forschungsgemeinschaft (DFG, German Research Foundation) -- Project ID 279384907, SFB 1245, and MA 4248/3-1.
M.R.W. and M.G. acknowledge support of the National Science and Technology Council, Taiwan under Grant No.~111-2628-M-001-003-MY4, and the Academia Sinica (Project Nos.~AS-CDA-109-M11 and AS-IV-114-M04).
M.R.W. also acknowledges support from the Physics Division of the National Center for Theoretical Sciences, Taiwan.
C.Y.L. acknowledges support from the National Center for High-performance Computing (NCHC).
We acknowledge the following software: \textsc{matplotlib}~\cite{matplotlib}, and \textsc{numpy}~\cite{numpy}.

The work is partially funded by the European Union. Views and opinions expressed are however those of the author(s) only and do not necessarily reflect those of the European Union or the European Research Council Executive Agency. Neither the European Union nor the granting authority can be held responsible for them.
\end{acknowledgments}

\bibliographystyle{apsrev4-1}
% \nocite{*}
%\bibliography{references.bib}

\begin{thebibliography}{85}%
\makeatletter
\providecommand \@ifxundefined [1]{%
 \@ifx{#1\undefined}
}%
\providecommand \@ifnum [1]{%
 \ifnum #1\expandafter \@firstoftwo
 \else \expandafter \@secondoftwo
 \fi
}%
\providecommand \@ifx [1]{%
 \ifx #1\expandafter \@firstoftwo
 \else \expandafter \@secondoftwo
 \fi
}%
\providecommand \natexlab [1]{#1}%
\providecommand \enquote  [1]{``#1''}%
\providecommand \bibnamefont  [1]{#1}%
\providecommand \bibfnamefont [1]{#1}%
\providecommand \citenamefont [1]{#1}%
\providecommand \href@noop [0]{\@secondoftwo}%
\providecommand \href [0]{\begingroup \@sanitize@url \@href}%
\providecommand \@href[1]{\@@startlink{#1}\@@href}%
\providecommand \@@href[1]{\endgroup#1\@@endlink}%
\providecommand \@sanitize@url [0]{\catcode `\\12\catcode `\$12\catcode
  `\&12\catcode `\#12\catcode `\^12\catcode `\_12\catcode `\%12\relax}%
\providecommand \@@startlink[1]{}%
\providecommand \@@endlink[0]{}%
\providecommand \url  [0]{\begingroup\@sanitize@url \@url }%
\providecommand \@url [1]{\endgroup\@href {#1}{\urlprefix }}%
\providecommand \urlprefix  [0]{URL }%
\providecommand \Eprint [0]{\href }%
\providecommand \doibase [0]{http://dx.doi.org/}%
\providecommand \selectlanguage [0]{\@gobble}%
\providecommand \bibinfo  [0]{\@secondoftwo}%
\providecommand \bibfield  [0]{\@secondoftwo}%
\providecommand \translation [1]{[#1]}%
\providecommand \BibitemOpen [0]{}%
\providecommand \bibitemStop [0]{}%
\providecommand \bibitemNoStop [0]{.\EOS\space}%
\providecommand \EOS [0]{\spacefactor3000\relax}%
\providecommand \BibitemShut  [1]{\csname bibitem#1\endcsname}%
\let\auto@bib@innerbib\@empty
%</preamble>
\bibitem [{\citenamefont {Janka}(2012)}]{janka2012explosion}%
  \BibitemOpen
  \bibfield  {author} {\bibinfo {author} {\bibfnamefont {H.-T.}\ \bibnamefont
  {Janka}},\ }\href {\doibase 10.1146/annurev-nucl-102711-094901} {\bibfield
  {journal} {\bibinfo  {journal} {Annu. Rev. Nucl. Part. Sci.}\ }\textbf
  {\bibinfo {volume} {62}},\ \bibinfo {pages} {407} (\bibinfo {year} {2012})},\
  \Eprint {http://arxiv.org/abs/1206.2503} {arXiv:1206.2503} \BibitemShut
  {NoStop}%
\bibitem [{\citenamefont {Pan}\ \emph {et~al.}(2019)\citenamefont {Pan},
  \citenamefont {Mattes}, \citenamefont {O'Connor}, \citenamefont {Couch},
  \citenamefont {Perego},\ and\ \citenamefont {Arcones}}]{Pan2018theimpact}%
  \BibitemOpen
  \bibfield  {author} {\bibinfo {author} {\bibfnamefont {K.-C.}\ \bibnamefont
  {Pan}}, \bibinfo {author} {\bibfnamefont {C.}~\bibnamefont {Mattes}},
  \bibinfo {author} {\bibfnamefont {E.~P.}\ \bibnamefont {O'Connor}}, \bibinfo
  {author} {\bibfnamefont {S.~M.}\ \bibnamefont {Couch}}, \bibinfo {author}
  {\bibfnamefont {A.}~\bibnamefont {Perego}}, \ and\ \bibinfo {author}
  {\bibfnamefont {A.}~\bibnamefont {Arcones}},\ }\href {\doibase
  10.1088/1361-6471/aaed51} {\bibfield  {journal} {\bibinfo  {journal} {J.
  Phys. G}\ }\textbf {\bibinfo {volume} {46}},\ \bibinfo {pages} {014001}
  (\bibinfo {year} {2019})},\ \Eprint {http://arxiv.org/abs/1806.10030}
  {arXiv:1806.10030 [astro-ph.HE]} \BibitemShut {NoStop}%
\bibitem [{\citenamefont {Burrows}\ \emph {et~al.}(2018)\citenamefont
  {Burrows}, \citenamefont {Vartanyan}, \citenamefont {Dolence}, \citenamefont
  {Skinner},\ and\ \citenamefont {Radice}}]{Burrows2016crucial}%
  \BibitemOpen
  \bibfield  {author} {\bibinfo {author} {\bibfnamefont {A.}~\bibnamefont
  {Burrows}}, \bibinfo {author} {\bibfnamefont {D.}~\bibnamefont {Vartanyan}},
  \bibinfo {author} {\bibfnamefont {J.~C.}\ \bibnamefont {Dolence}}, \bibinfo
  {author} {\bibfnamefont {M.~A.}\ \bibnamefont {Skinner}}, \ and\ \bibinfo
  {author} {\bibfnamefont {D.}~\bibnamefont {Radice}},\ }\href {\doibase
  10.1007/s11214-017-0450-9} {\bibfield  {journal} {\bibinfo  {journal} {Space
  Sci. Rev.}\ }\textbf {\bibinfo {volume} {214}},\ \bibinfo {pages} {33}
  (\bibinfo {year} {2018})},\ \Eprint {http://arxiv.org/abs/1611.05859}
  {arXiv:1611.05859 [astro-ph.SR]} \BibitemShut {NoStop}%
\bibitem [{\citenamefont {Richers}\ \emph {et~al.}(2017)\citenamefont
  {Richers}, \citenamefont {Nagakura}, \citenamefont {Ott}, \citenamefont
  {Dolence}, \citenamefont {Sumiyoshi},\ and\ \citenamefont
  {Yamada}}]{Richers2017detailed}%
  \BibitemOpen
  \bibfield  {author} {\bibinfo {author} {\bibfnamefont {S.}~\bibnamefont
  {Richers}}, \bibinfo {author} {\bibfnamefont {H.}~\bibnamefont {Nagakura}},
  \bibinfo {author} {\bibfnamefont {C.~D.}\ \bibnamefont {Ott}}, \bibinfo
  {author} {\bibfnamefont {J.}~\bibnamefont {Dolence}}, \bibinfo {author}
  {\bibfnamefont {K.}~\bibnamefont {Sumiyoshi}}, \ and\ \bibinfo {author}
  {\bibfnamefont {S.}~\bibnamefont {Yamada}},\ }\href {\doibase
  10.3847/1538-4357/aa8bb2} {\bibfield  {journal} {\bibinfo  {journal}
  {Astrophys. J.}\ }\textbf {\bibinfo {volume} {847}},\ \bibinfo {pages} {133}
  (\bibinfo {year} {2017})},\ \Eprint {http://arxiv.org/abs/1706.06187}
  {arXiv:1706.06187 [astro-ph.HE]} \BibitemShut {NoStop}%
\bibitem [{\citenamefont {Nagakura}\ \emph {et~al.}(2018)\citenamefont
  {Nagakura}, \citenamefont {Iwakami}, \citenamefont {Furusawa}, \citenamefont
  {Okawa}, \citenamefont {Harada}, \citenamefont {Sumiyoshi}, \citenamefont
  {Yamada}, \citenamefont {Matsufuru},\ and\ \citenamefont
  {Imakura}}]{Nagakura:2017mnp}%
  \BibitemOpen
  \bibfield  {author} {\bibinfo {author} {\bibfnamefont {H.}~\bibnamefont
  {Nagakura}}, \bibinfo {author} {\bibfnamefont {W.}~\bibnamefont {Iwakami}},
  \bibinfo {author} {\bibfnamefont {S.}~\bibnamefont {Furusawa}}, \bibinfo
  {author} {\bibfnamefont {H.}~\bibnamefont {Okawa}}, \bibinfo {author}
  {\bibfnamefont {A.}~\bibnamefont {Harada}}, \bibinfo {author} {\bibfnamefont
  {K.}~\bibnamefont {Sumiyoshi}}, \bibinfo {author} {\bibfnamefont
  {S.}~\bibnamefont {Yamada}}, \bibinfo {author} {\bibfnamefont
  {H.}~\bibnamefont {Matsufuru}}, \ and\ \bibinfo {author} {\bibfnamefont
  {A.}~\bibnamefont {Imakura}},\ }\href {\doibase 10.3847/1538-4357/aaac29}
  {\bibfield  {journal} {\bibinfo  {journal} {Astrophys. J.}\ }\textbf
  {\bibinfo {volume} {854}},\ \bibinfo {pages} {136} (\bibinfo {year}
  {2018})},\ \Eprint {http://arxiv.org/abs/1702.01752} {arXiv:1702.01752
  [astro-ph.HE]} \BibitemShut {NoStop}%
\bibitem [{\citenamefont {{O'Connor}}\ \emph {et~al.}(2018)\citenamefont
  {{O'Connor}}, \citenamefont {{Bollig}}, \citenamefont {{Burrows}},
  \citenamefont {{Couch}}, \citenamefont {{Fischer}}, \citenamefont {{Janka}},
  \citenamefont {{Kotake}}, \citenamefont {{Lentz}}, \citenamefont
  {{Liebend{\"o}rfer}}, \citenamefont {{Messer}}, \citenamefont {{Mezzacappa}},
  \citenamefont {{Takiwaki}},\ and\ \citenamefont
  {{Vartanyan}}}]{oconnor2018global}%
  \BibitemOpen
  \bibfield  {author} {\bibinfo {author} {\bibfnamefont {E.}~\bibnamefont
  {{O'Connor}}}, \bibinfo {author} {\bibfnamefont {R.}~\bibnamefont
  {{Bollig}}}, \bibinfo {author} {\bibfnamefont {A.}~\bibnamefont {{Burrows}}},
  \bibinfo {author} {\bibfnamefont {S.}~\bibnamefont {{Couch}}}, \bibinfo
  {author} {\bibfnamefont {T.}~\bibnamefont {{Fischer}}}, \bibinfo {author}
  {\bibfnamefont {H.-T.}\ \bibnamefont {{Janka}}}, \bibinfo {author}
  {\bibfnamefont {K.}~\bibnamefont {{Kotake}}}, \bibinfo {author}
  {\bibfnamefont {E.~J.}\ \bibnamefont {{Lentz}}}, \bibinfo {author}
  {\bibfnamefont {M.}~\bibnamefont {{Liebend{\"o}rfer}}}, \bibinfo {author}
  {\bibfnamefont {O.~E.~B.}\ \bibnamefont {{Messer}}}, \bibinfo {author}
  {\bibfnamefont {A.}~\bibnamefont {{Mezzacappa}}}, \bibinfo {author}
  {\bibfnamefont {T.}~\bibnamefont {{Takiwaki}}}, \ and\ \bibinfo {author}
  {\bibfnamefont {D.}~\bibnamefont {{Vartanyan}}},\ }\href {\doibase
  10.1088/1361-6471/aadeae} {\bibfield  {journal} {\bibinfo  {journal} {Journal
  of Physics G Nuclear Physics}\ }\textbf {\bibinfo {volume} {45}},\ \bibinfo
  {pages} {104001} (\bibinfo {year} {2018})},\ \Eprint
  {http://arxiv.org/abs/1806.04175} {arXiv:1806.04175 [astro-ph.HE]}
  \BibitemShut {NoStop}%
\bibitem [{\citenamefont {Mezzacappa}\ \emph {et~al.}(2020)\citenamefont
  {Mezzacappa}, \citenamefont {Endeve}, \citenamefont {Bronson~Messer},\ and\
  \citenamefont {Bruenn}}]{Mezzacappa2020physical}%
  \BibitemOpen
  \bibfield  {author} {\bibinfo {author} {\bibfnamefont {A.}~\bibnamefont
  {Mezzacappa}}, \bibinfo {author} {\bibfnamefont {E.}~\bibnamefont {Endeve}},
  \bibinfo {author} {\bibfnamefont {O.~E.}\ \bibnamefont {Bronson~Messer}}, \
  and\ \bibinfo {author} {\bibfnamefont {S.~W.}\ \bibnamefont {Bruenn}},\
  }\href {\doibase 10.1007/s41115-020-00010-8} {\bibfield  {journal} {\bibinfo
  {journal} {Liv. Rev. Comput. Astrophys.}\ }\textbf {\bibinfo {volume} {6}},\
  \bibinfo {pages} {4} (\bibinfo {year} {2020})},\ \Eprint
  {http://arxiv.org/abs/2010.09013} {arXiv:2010.09013 [astro-ph.HE]}
  \BibitemShut {NoStop}%
\bibitem [{\citenamefont {Foucart}(2023)}]{Foucart2022neutrino}%
  \BibitemOpen
  \bibfield  {author} {\bibinfo {author} {\bibfnamefont {F.}~\bibnamefont
  {Foucart}},\ }\href {\doibase 10.1007/s41115-023-00016-y} {\bibfield
  {journal} {\bibinfo  {journal} {Liv. Rev. Comput. Astrophys.}\ }\textbf
  {\bibinfo {volume} {9}},\ \bibinfo {pages} {1} (\bibinfo {year} {2023})},\
  \Eprint {http://arxiv.org/abs/2209.02538} {arXiv:2209.02538 [astro-ph.HE]}
  \BibitemShut {NoStop}%
\bibitem [{\citenamefont {Capozzi}\ and\ \citenamefont
  {Saviano}(2022)}]{capozzi2022neutrino}%
  \BibitemOpen
  \bibfield  {author} {\bibinfo {author} {\bibfnamefont {F.}~\bibnamefont
  {Capozzi}}\ and\ \bibinfo {author} {\bibfnamefont {N.}~\bibnamefont
  {Saviano}},\ }\href {\doibase 10.3390/universe8020094} {\bibfield  {journal}
  {\bibinfo  {journal} {Universe}\ }\textbf {\bibinfo {volume} {8}},\ \bibinfo
  {pages} {94} (\bibinfo {year} {2022})},\ \Eprint
  {http://arxiv.org/abs/2202.02494} {arXiv:2202.02494} \BibitemShut {NoStop}%
\bibitem [{\citenamefont {Volpe}(2024)}]{volpe2023neutrinos}%
  \BibitemOpen
  \bibfield  {author} {\bibinfo {author} {\bibfnamefont {M.~C.}\ \bibnamefont
  {Volpe}},\ }\href {\doibase 10.1103/RevModPhys.96.025004} {\bibfield
  {journal} {\bibinfo  {journal} {Rev. Mod. Phys.}\ }\textbf {\bibinfo {volume}
  {96}},\ \bibinfo {pages} {025004} (\bibinfo {year} {2024})},\ \Eprint
  {http://arxiv.org/abs/2301.11814} {arXiv:2301.11814 [hep-ph]} \BibitemShut
  {NoStop}%
\bibitem [{\citenamefont {Fischer}\ \emph {et~al.}(2024)\citenamefont
  {Fischer}, \citenamefont {Guo}, \citenamefont {Langanke}, \citenamefont
  {Martinez-Pinedo}, \citenamefont {Qian},\ and\ \citenamefont
  {Wu}}]{Fischer:2023ebq}%
  \BibitemOpen
  \bibfield  {author} {\bibinfo {author} {\bibfnamefont {T.}~\bibnamefont
  {Fischer}}, \bibinfo {author} {\bibfnamefont {G.}~\bibnamefont {Guo}},
  \bibinfo {author} {\bibfnamefont {K.}~\bibnamefont {Langanke}}, \bibinfo
  {author} {\bibfnamefont {G.}~\bibnamefont {Martinez-Pinedo}}, \bibinfo
  {author} {\bibfnamefont {Y.-Z.}\ \bibnamefont {Qian}}, \ and\ \bibinfo
  {author} {\bibfnamefont {M.-R.}\ \bibnamefont {Wu}},\ }\href {\doibase
  10.1016/j.ppnp.2024.104107} {\bibfield  {journal} {\bibinfo  {journal} {Prog.
  Part. Nucl. Phys.}\ }\textbf {\bibinfo {volume} {137}},\ \bibinfo {pages}
  {104107} (\bibinfo {year} {2024})},\ \Eprint
  {http://arxiv.org/abs/2308.03962} {arXiv:2308.03962 [astro-ph.HE]}
  \BibitemShut {NoStop}%
\bibitem [{\citenamefont {{Sigl}}\ and\ \citenamefont
  {{Raffelt}}(1993)}]{sigl1993general}%
  \BibitemOpen
  \bibfield  {author} {\bibinfo {author} {\bibfnamefont {G.}~\bibnamefont
  {{Sigl}}}\ and\ \bibinfo {author} {\bibfnamefont {G.}~\bibnamefont
  {{Raffelt}}},\ }\href {\doibase 10.1016/0550-3213(93)90175-O} {\bibfield
  {journal} {\bibinfo  {journal} {Nuclear Physics B}\ }\textbf {\bibinfo
  {volume} {406}},\ \bibinfo {pages} {423} (\bibinfo {year}
  {1993})}\BibitemShut {NoStop}%
\bibitem [{\citenamefont {Vlasenko}\ \emph {et~al.}(2014)\citenamefont
  {Vlasenko}, \citenamefont {Fuller},\ and\ \citenamefont
  {Cirigliano}}]{vlasenko2014neutrino}%
  \BibitemOpen
  \bibfield  {author} {\bibinfo {author} {\bibfnamefont {A.}~\bibnamefont
  {Vlasenko}}, \bibinfo {author} {\bibfnamefont {G.~M.}\ \bibnamefont
  {Fuller}}, \ and\ \bibinfo {author} {\bibfnamefont {V.}~\bibnamefont
  {Cirigliano}},\ }\href {\doibase 10.1103/PhysRevD.89.105004} {\bibfield
  {journal} {\bibinfo  {journal} {Phys. Rev. D}\ }\textbf {\bibinfo {volume}
  {89}},\ \bibinfo {pages} {105004} (\bibinfo {year} {2014})},\ \Eprint
  {http://arxiv.org/abs/1309.2628} {arXiv:1309.2628} \BibitemShut {NoStop}%
\bibitem [{\citenamefont {Volpe}(2015)}]{volpe2015neutrino}%
  \BibitemOpen
  \bibfield  {author} {\bibinfo {author} {\bibfnamefont {C.}~\bibnamefont
  {Volpe}},\ }\href {\doibase 10.1142/S0218301315410098} {\bibfield  {journal}
  {\bibinfo  {journal} {Int. J. Mod. Phys. E}\ }\textbf {\bibinfo {volume}
  {24}},\ \bibinfo {pages} {1541009} (\bibinfo {year} {2015})},\ \Eprint
  {http://arxiv.org/abs/1506.06222} {arXiv:1506.06222} \BibitemShut {NoStop}%
\bibitem [{\citenamefont {Kartavtsev}\ \emph {et~al.}(2015)\citenamefont
  {Kartavtsev}, \citenamefont {Raffelt},\ and\ \citenamefont
  {Vogel}}]{Kartavtsev2015neutrino}%
  \BibitemOpen
  \bibfield  {author} {\bibinfo {author} {\bibfnamefont {A.}~\bibnamefont
  {Kartavtsev}}, \bibinfo {author} {\bibfnamefont {G.}~\bibnamefont {Raffelt}},
  \ and\ \bibinfo {author} {\bibfnamefont {H.}~\bibnamefont {Vogel}},\ }\href
  {\doibase 10.1103/PhysRevD.91.125020} {\bibfield  {journal} {\bibinfo
  {journal} {Phys. Rev. D}\ }\textbf {\bibinfo {volume} {91}},\ \bibinfo
  {pages} {125020} (\bibinfo {year} {2015})},\ \Eprint
  {http://arxiv.org/abs/1504.03230} {arXiv:1504.03230 [hep-ph]} \BibitemShut
  {NoStop}%
\bibitem [{\citenamefont {{Blaschke}}\ and\ \citenamefont
  {{Cirigliano}}(2016)}]{blaschke2016neutrino}%
  \BibitemOpen
  \bibfield  {author} {\bibinfo {author} {\bibfnamefont {D.~N.}\ \bibnamefont
  {{Blaschke}}}\ and\ \bibinfo {author} {\bibfnamefont {V.}~\bibnamefont
  {{Cirigliano}}},\ }\href {\doibase 10.1103/PhysRevD.94.033009} {\bibfield
  {journal} {\bibinfo  {journal} {Phys. Rev. D}\ }\textbf {\bibinfo {volume}
  {94}},\ \bibinfo {eid} {033009} (\bibinfo {year} {2016})},\ \Eprint
  {http://arxiv.org/abs/1605.09383} {arXiv:1605.09383 [hep-ph]} \BibitemShut
  {NoStop}%
\bibitem [{\citenamefont {Richers}\ \emph {et~al.}(2019)\citenamefont
  {Richers}, \citenamefont {McLaughlin}, \citenamefont {Kneller},\ and\
  \citenamefont {Vlasenko}}]{Richers2019neutrino}%
  \BibitemOpen
  \bibfield  {author} {\bibinfo {author} {\bibfnamefont {S.~A.}\ \bibnamefont
  {Richers}}, \bibinfo {author} {\bibfnamefont {G.~C.}\ \bibnamefont
  {McLaughlin}}, \bibinfo {author} {\bibfnamefont {J.~P.}\ \bibnamefont
  {Kneller}}, \ and\ \bibinfo {author} {\bibfnamefont {A.}~\bibnamefont
  {Vlasenko}},\ }\href {\doibase 10.1103/PhysRevD.99.123014} {\bibfield
  {journal} {\bibinfo  {journal} {Phys. Rev. D}\ }\textbf {\bibinfo {volume}
  {99}},\ \bibinfo {pages} {123014} (\bibinfo {year} {2019})},\ \Eprint
  {http://arxiv.org/abs/1903.00022} {arXiv:1903.00022 [astro-ph.HE]}
  \BibitemShut {NoStop}%
\bibitem [{\citenamefont {Fiorillo}\ \emph {et~al.}(2024)\citenamefont
  {Fiorillo}, \citenamefont {Raffelt},\ and\ \citenamefont
  {Sigl}}]{fiorillo2024inhomogenous}%
  \BibitemOpen
  \bibfield  {author} {\bibinfo {author} {\bibfnamefont {D.~F.~G.}\
  \bibnamefont {Fiorillo}}, \bibinfo {author} {\bibfnamefont {G.~G.}\
  \bibnamefont {Raffelt}}, \ and\ \bibinfo {author} {\bibfnamefont
  {G.}~\bibnamefont {Sigl}},\ }\href {\doibase 10.1103/PhysRevLett.133.021002}
  {\bibfield  {journal} {\bibinfo  {journal} {Phys. Rev. Lett.}\ }\textbf
  {\bibinfo {volume} {133}},\ \bibinfo {pages} {021002} (\bibinfo {year}
  {2024})},\ \Eprint {http://arxiv.org/abs/2401.05278} {arXiv:2401.05278
  [hep-ph]} \BibitemShut {NoStop}%
\bibitem [{\citenamefont {Sawyer}(2009)}]{sawyer2009multiangle}%
  \BibitemOpen
  \bibfield  {author} {\bibinfo {author} {\bibfnamefont {R.~F.}\ \bibnamefont
  {Sawyer}},\ }\href {\doibase 10.1103/PhysRevD.79.105003} {\bibfield
  {journal} {\bibinfo  {journal} {Phys. Rev. D}\ }\textbf {\bibinfo {volume}
  {79}},\ \bibinfo {pages} {105003} (\bibinfo {year} {2009})},\ \Eprint
  {http://arxiv.org/abs/arXiv:0803.4319} {arXiv:arXiv:0803.4319} \BibitemShut
  {NoStop}%
\bibitem [{\citenamefont {{Izaguirre}}\ \emph {et~al.}(2017)\citenamefont
  {{Izaguirre}}, \citenamefont {{Raffelt}},\ and\ \citenamefont
  {{Tamborra}}}]{izaguirre2017fast}%
  \BibitemOpen
  \bibfield  {author} {\bibinfo {author} {\bibfnamefont {I.}~\bibnamefont
  {{Izaguirre}}}, \bibinfo {author} {\bibfnamefont {G.}~\bibnamefont
  {{Raffelt}}}, \ and\ \bibinfo {author} {\bibfnamefont {I.}~\bibnamefont
  {{Tamborra}}},\ }\href {\doibase 10.1103/PhysRevLett.118.021101} {\bibfield
  {journal} {\bibinfo  {journal} {Phys. Rev. Lett.}\ }\textbf {\bibinfo
  {volume} {118}},\ \bibinfo {eid} {021101} (\bibinfo {year} {2017})},\ \Eprint
  {http://arxiv.org/abs/1610.01612} {arXiv:1610.01612 [hep-ph]} \BibitemShut
  {NoStop}%
\bibitem [{\citenamefont {Capozzi}\ \emph {et~al.}(2017)\citenamefont
  {Capozzi}, \citenamefont {Dasgupta}, \citenamefont {Lisi}, \citenamefont
  {Marrone},\ and\ \citenamefont {Mirizzi}}]{capozzi2017fast}%
  \BibitemOpen
  \bibfield  {author} {\bibinfo {author} {\bibfnamefont {F.}~\bibnamefont
  {Capozzi}}, \bibinfo {author} {\bibfnamefont {B.}~\bibnamefont {Dasgupta}},
  \bibinfo {author} {\bibfnamefont {E.}~\bibnamefont {Lisi}}, \bibinfo {author}
  {\bibfnamefont {A.}~\bibnamefont {Marrone}}, \ and\ \bibinfo {author}
  {\bibfnamefont {A.}~\bibnamefont {Mirizzi}},\ }\href {\doibase
  10.1103/PhysRevD.96.043016} {\bibfield  {journal} {\bibinfo  {journal} {Phys.
  Rev. D}\ }\textbf {\bibinfo {volume} {96}},\ \bibinfo {pages} {043016}
  (\bibinfo {year} {2017})},\ \Eprint {http://arxiv.org/abs/1706.03360}
  {arXiv:1706.03360} \BibitemShut {NoStop}%
\bibitem [{\citenamefont {{Yi}}\ \emph {et~al.}(2019)\citenamefont {{Yi}},
  \citenamefont {{Ma}}, \citenamefont {{Martin}},\ and\ \citenamefont
  {{Duan}}}]{yi2019dispersion}%
  \BibitemOpen
  \bibfield  {author} {\bibinfo {author} {\bibfnamefont {C.}~\bibnamefont
  {{Yi}}}, \bibinfo {author} {\bibfnamefont {L.}~\bibnamefont {{Ma}}}, \bibinfo
  {author} {\bibfnamefont {J.~D.}\ \bibnamefont {{Martin}}}, \ and\ \bibinfo
  {author} {\bibfnamefont {H.}~\bibnamefont {{Duan}}},\ }\href {\doibase
  10.1103/PhysRevD.99.063005} {\bibfield  {journal} {\bibinfo  {journal} {Phys.
  Rev. D}\ }\textbf {\bibinfo {volume} {99}},\ \bibinfo {eid} {063005}
  (\bibinfo {year} {2019})},\ \Eprint {http://arxiv.org/abs/1901.01546}
  {arXiv:1901.01546 [hep-ph]} \BibitemShut {NoStop}%
\bibitem [{\citenamefont {Morinaga}(2022)}]{morinaga2022fast}%
  \BibitemOpen
  \bibfield  {author} {\bibinfo {author} {\bibfnamefont {T.}~\bibnamefont
  {Morinaga}},\ }\href {\doibase 10.1103/PhysRevD.105.L101301} {\bibfield
  {journal} {\bibinfo  {journal} {Phys. Rev. D}\ }\textbf {\bibinfo {volume}
  {105}},\ \bibinfo {pages} {L101301} (\bibinfo {year} {2022})},\ \Eprint
  {http://arxiv.org/abs/2103.15267} {arXiv:2103.15267} \BibitemShut {NoStop}%
\bibitem [{\citenamefont {Dasgupta}(2022)}]{dasgupta2022collective}%
  \BibitemOpen
  \bibfield  {author} {\bibinfo {author} {\bibfnamefont {B.}~\bibnamefont
  {Dasgupta}},\ }\href {\doibase 10.1103/PhysRevLett.128.081102} {\bibfield
  {journal} {\bibinfo  {journal} {Phys. Rev. Lett.}\ }\textbf {\bibinfo
  {volume} {128}},\ \bibinfo {pages} {081102} (\bibinfo {year} {2022})},\
  \Eprint {http://arxiv.org/abs/2110.00192} {arXiv:2110.00192} \BibitemShut
  {NoStop}%
\bibitem [{\citenamefont {Wu}\ and\ \citenamefont
  {Tamborra}(2017)}]{wu2017fast}%
  \BibitemOpen
  \bibfield  {author} {\bibinfo {author} {\bibfnamefont {M.~R.}\ \bibnamefont
  {Wu}}\ and\ \bibinfo {author} {\bibfnamefont {I.}~\bibnamefont {Tamborra}},\
  }\href {\doibase 10.1103/PhysRevD.95.103007} {\bibfield  {journal} {\bibinfo
  {journal} {Phys. Rev. D}\ }\textbf {\bibinfo {volume} {95}},\ \bibinfo
  {pages} {103007} (\bibinfo {year} {2017})}\BibitemShut {NoStop}%
\bibitem [{\citenamefont {{Wu}}\ \emph {et~al.}(2017)\citenamefont {{Wu}},
  \citenamefont {{Tamborra}}, \citenamefont {{Just}},\ and\ \citenamefont
  {{Janka}}}]{wu2017imprints}%
  \BibitemOpen
  \bibfield  {author} {\bibinfo {author} {\bibfnamefont {M.-R.}\ \bibnamefont
  {{Wu}}}, \bibinfo {author} {\bibfnamefont {I.}~\bibnamefont {{Tamborra}}},
  \bibinfo {author} {\bibfnamefont {O.}~\bibnamefont {{Just}}}, \ and\ \bibinfo
  {author} {\bibfnamefont {H.-T.}\ \bibnamefont {{Janka}}},\ }\href {\doibase
  10.1103/PhysRevD.96.123015} {\bibfield  {journal} {\bibinfo  {journal} {Phys.
  Rev. D}\ }\textbf {\bibinfo {volume} {96}},\ \bibinfo {eid} {123015}
  (\bibinfo {year} {2017})},\ \Eprint {http://arxiv.org/abs/1711.00477}
  {arXiv:1711.00477 [astro-ph.HE]} \BibitemShut {NoStop}%
\bibitem [{\citenamefont {{Delfan Azari}}\ \emph {et~al.}(2019)\citenamefont
  {{Delfan Azari}}, \citenamefont {Yamada}, \citenamefont {Morinaga},
  \citenamefont {Iwakami}, \citenamefont {Okawa}, \citenamefont {Nagakura},\
  and\ \citenamefont {Sumiyoshi}}]{azari2019linear}%
  \BibitemOpen
  \bibfield  {author} {\bibinfo {author} {\bibfnamefont {M.}~\bibnamefont
  {{Delfan Azari}}}, \bibinfo {author} {\bibfnamefont {S.}~\bibnamefont
  {Yamada}}, \bibinfo {author} {\bibfnamefont {T.}~\bibnamefont {Morinaga}},
  \bibinfo {author} {\bibfnamefont {W.}~\bibnamefont {Iwakami}}, \bibinfo
  {author} {\bibfnamefont {H.}~\bibnamefont {Okawa}}, \bibinfo {author}
  {\bibfnamefont {H.}~\bibnamefont {Nagakura}}, \ and\ \bibinfo {author}
  {\bibfnamefont {K.}~\bibnamefont {Sumiyoshi}},\ }\href {\doibase
  10.1103/PhysRevD.99.103011} {\bibfield  {journal} {\bibinfo  {journal} {Phys.
  Rev. D}\ }\textbf {\bibinfo {volume} {99}},\ \bibinfo {pages} {103011}
  (\bibinfo {year} {2019})},\ \Eprint {http://arxiv.org/abs/1902.07467}
  {arXiv:1902.07467} \BibitemShut {NoStop}%
\bibitem [{\citenamefont {{Delfan Azari}}\ \emph {et~al.}(2020)\citenamefont
  {{Delfan Azari}}, \citenamefont {{Yamada}}, \citenamefont {{Morinaga}},
  \citenamefont {{Nagakura}}, \citenamefont {{Furusawa}}, \citenamefont
  {{Harada}}, \citenamefont {{Okawa}}, \citenamefont {{Iwakami}},\ and\
  \citenamefont {{Sumiyoshi}}}]{azari2020fast}%
  \BibitemOpen
  \bibfield  {author} {\bibinfo {author} {\bibfnamefont {M.}~\bibnamefont
  {{Delfan Azari}}}, \bibinfo {author} {\bibfnamefont {S.}~\bibnamefont
  {{Yamada}}}, \bibinfo {author} {\bibfnamefont {T.}~\bibnamefont
  {{Morinaga}}}, \bibinfo {author} {\bibfnamefont {H.}~\bibnamefont
  {{Nagakura}}}, \bibinfo {author} {\bibfnamefont {S.}~\bibnamefont
  {{Furusawa}}}, \bibinfo {author} {\bibfnamefont {A.}~\bibnamefont
  {{Harada}}}, \bibinfo {author} {\bibfnamefont {H.}~\bibnamefont {{Okawa}}},
  \bibinfo {author} {\bibfnamefont {W.}~\bibnamefont {{Iwakami}}}, \ and\
  \bibinfo {author} {\bibfnamefont {K.}~\bibnamefont {{Sumiyoshi}}},\ }\href
  {\doibase 10.1103/PhysRevD.101.023018} {\bibfield  {journal} {\bibinfo
  {journal} {Phys. Rev. D}\ }\textbf {\bibinfo {volume} {101}},\ \bibinfo {eid}
  {023018} (\bibinfo {year} {2020})},\ \Eprint
  {http://arxiv.org/abs/1910.06176} {arXiv:1910.06176 [astro-ph.HE]}
  \BibitemShut {NoStop}%
\bibitem [{\citenamefont {Nagakura}\ \emph {et~al.}(2019)\citenamefont
  {Nagakura}, \citenamefont {Morinaga}, \citenamefont {Kato},\ and\
  \citenamefont {Yamada}}]{nagakura2019fast}%
  \BibitemOpen
  \bibfield  {author} {\bibinfo {author} {\bibfnamefont {H.}~\bibnamefont
  {Nagakura}}, \bibinfo {author} {\bibfnamefont {T.}~\bibnamefont {Morinaga}},
  \bibinfo {author} {\bibfnamefont {C.}~\bibnamefont {Kato}}, \ and\ \bibinfo
  {author} {\bibfnamefont {S.}~\bibnamefont {Yamada}},\ }\href {\doibase
  10.3847/1538-4357/ab4cf2} {\bibfield  {journal} {\bibinfo  {journal}
  {Astrophys. J.}\ }\textbf {\bibinfo {volume} {886}},\ \bibinfo {pages} {139}
  (\bibinfo {year} {2019})},\ \Eprint {http://arxiv.org/abs/1910.04288}
  {arXiv:1910.04288} \BibitemShut {NoStop}%
\bibitem [{\citenamefont {{Morinaga}}\ \emph {et~al.}(2020)\citenamefont
  {{Morinaga}}, \citenamefont {{Nagakura}}, \citenamefont {{Kato}},\ and\
  \citenamefont {{Yamada}}}]{morinaga2020fast}%
  \BibitemOpen
  \bibfield  {author} {\bibinfo {author} {\bibfnamefont {T.}~\bibnamefont
  {{Morinaga}}}, \bibinfo {author} {\bibfnamefont {H.}~\bibnamefont
  {{Nagakura}}}, \bibinfo {author} {\bibfnamefont {C.}~\bibnamefont {{Kato}}},
  \ and\ \bibinfo {author} {\bibfnamefont {S.}~\bibnamefont {{Yamada}}},\
  }\href {\doibase 10.1103/PhysRevResearch.2.012046} {\bibfield  {journal}
  {\bibinfo  {journal} {Phys. Rev. Res.}\ }\textbf {\bibinfo {volume} {2}},\
  \bibinfo {eid} {012046(R)} (\bibinfo {year} {2020})},\ \Eprint
  {http://arxiv.org/abs/1909.13131} {arXiv:1909.13131 [astro-ph.HE]}
  \BibitemShut {NoStop}%
\bibitem [{\citenamefont {Abbar}\ \emph {et~al.}(2020)\citenamefont {Abbar},
  \citenamefont {Duan}, \citenamefont {Sumiyoshi}, \citenamefont {Takiwaki},\
  and\ \citenamefont {Volpe}}]{abbar2020fast}%
  \BibitemOpen
  \bibfield  {author} {\bibinfo {author} {\bibfnamefont {S.}~\bibnamefont
  {Abbar}}, \bibinfo {author} {\bibfnamefont {H.}~\bibnamefont {Duan}},
  \bibinfo {author} {\bibfnamefont {K.}~\bibnamefont {Sumiyoshi}}, \bibinfo
  {author} {\bibfnamefont {T.}~\bibnamefont {Takiwaki}}, \ and\ \bibinfo
  {author} {\bibfnamefont {M.~C.}\ \bibnamefont {Volpe}},\ }\href {\doibase
  10.1103/PhysRevD.101.043016} {\bibfield  {journal} {\bibinfo  {journal}
  {Phys. Rev. D}\ }\textbf {\bibinfo {volume} {101}},\ \bibinfo {pages}
  {043016} (\bibinfo {year} {2020})},\ \Eprint
  {http://arxiv.org/abs/1911.01983} {arXiv:1911.01983} \BibitemShut {NoStop}%
\bibitem [{\citenamefont {{George}}\ \emph {et~al.}(2020)\citenamefont
  {{George}}, \citenamefont {{Wu}}, \citenamefont {{Tamborra}}, \citenamefont
  {{Ardevol-Pulpillo}},\ and\ \citenamefont {{Janka}}}]{george2020fast}%
  \BibitemOpen
  \bibfield  {author} {\bibinfo {author} {\bibfnamefont {M.}~\bibnamefont
  {{George}}}, \bibinfo {author} {\bibfnamefont {M.-R.}\ \bibnamefont {{Wu}}},
  \bibinfo {author} {\bibfnamefont {I.}~\bibnamefont {{Tamborra}}}, \bibinfo
  {author} {\bibfnamefont {R.}~\bibnamefont {{Ardevol-Pulpillo}}}, \ and\
  \bibinfo {author} {\bibfnamefont {H.-T.}\ \bibnamefont {{Janka}}},\ }\href
  {\doibase 10.1103/PhysRevD.102.103015} {\bibfield  {journal} {\bibinfo
  {journal} {Phys. Rev. D}\ }\textbf {\bibinfo {volume} {102}},\ \bibinfo {eid}
  {103015} (\bibinfo {year} {2020})},\ \Eprint
  {http://arxiv.org/abs/2009.04046} {arXiv:2009.04046 [astro-ph.HE]}
  \BibitemShut {NoStop}%
\bibitem [{\citenamefont {Glas}\ \emph {et~al.}(2020)\citenamefont {Glas},
  \citenamefont {Janka}, \citenamefont {Capozzi}, \citenamefont {Sen},
  \citenamefont {Dasgupta}, \citenamefont {Mirizzi},\ and\ \citenamefont
  {Sigl}}]{glas2020fast}%
  \BibitemOpen
  \bibfield  {author} {\bibinfo {author} {\bibfnamefont {R.}~\bibnamefont
  {Glas}}, \bibinfo {author} {\bibfnamefont {H.-T.}\ \bibnamefont {Janka}},
  \bibinfo {author} {\bibfnamefont {F.}~\bibnamefont {Capozzi}}, \bibinfo
  {author} {\bibfnamefont {M.}~\bibnamefont {Sen}}, \bibinfo {author}
  {\bibfnamefont {B.}~\bibnamefont {Dasgupta}}, \bibinfo {author}
  {\bibfnamefont {A.}~\bibnamefont {Mirizzi}}, \ and\ \bibinfo {author}
  {\bibfnamefont {G.}~\bibnamefont {Sigl}},\ }\href {\doibase
  10.1103/physrevd.101.063001} {\bibfield  {journal} {\bibinfo  {journal}
  {Phys. Rev. D}\ }\textbf {\bibinfo {volume} {101}},\ \bibinfo {pages}
  {063001} (\bibinfo {year} {2020})},\ \Eprint
  {http://arxiv.org/abs/1912.00274} {arXiv:1912.00274} \BibitemShut {NoStop}%
\bibitem [{\citenamefont {{Nagakura}}\ \emph {et~al.}(2021)\citenamefont
  {{Nagakura}}, \citenamefont {{Burrows}}, \citenamefont {{Johns}},\ and\
  \citenamefont {{Fuller}}}]{nagakura2021occurrence}%
  \BibitemOpen
  \bibfield  {author} {\bibinfo {author} {\bibfnamefont {H.}~\bibnamefont
  {{Nagakura}}}, \bibinfo {author} {\bibfnamefont {A.}~\bibnamefont
  {{Burrows}}}, \bibinfo {author} {\bibfnamefont {L.}~\bibnamefont {{Johns}}},
  \ and\ \bibinfo {author} {\bibfnamefont {G.~M.}\ \bibnamefont {{Fuller}}},\
  }\href {\doibase 10.1103/PhysRevD.104.083025} {\bibfield  {journal} {\bibinfo
   {journal} {Phys. Rev. D}\ }\textbf {\bibinfo {volume} {104}},\ \bibinfo
  {eid} {083025} (\bibinfo {year} {2021})},\ \Eprint
  {http://arxiv.org/abs/2108.07281} {arXiv:2108.07281 [astro-ph.HE]}
  \BibitemShut {NoStop}%
\bibitem [{\citenamefont {Abbar}\ \emph {et~al.}(2021)\citenamefont {Abbar},
  \citenamefont {Capozzi}, \citenamefont {Glas}, \citenamefont {Janka},\ and\
  \citenamefont {Tamborra}}]{abbar2021characteristics}%
  \BibitemOpen
  \bibfield  {author} {\bibinfo {author} {\bibfnamefont {S.}~\bibnamefont
  {Abbar}}, \bibinfo {author} {\bibfnamefont {F.}~\bibnamefont {Capozzi}},
  \bibinfo {author} {\bibfnamefont {R.}~\bibnamefont {Glas}}, \bibinfo {author}
  {\bibfnamefont {H.~T.}\ \bibnamefont {Janka}}, \ and\ \bibinfo {author}
  {\bibfnamefont {I.}~\bibnamefont {Tamborra}},\ }\href {\doibase
  10.1103/PhysRevD.103.063033} {\bibfield  {journal} {\bibinfo  {journal}
  {Phys. Rev. D}\ }\textbf {\bibinfo {volume} {103}},\ \bibinfo {pages}
  {063033} (\bibinfo {year} {2021})},\ \Eprint
  {http://arxiv.org/abs/2012.06594} {arXiv:2012.06594} \BibitemShut {NoStop}%
\bibitem [{\citenamefont {{Harada}}\ and\ \citenamefont
  {{Nagakura}}(2022)}]{harada2022prospects}%
  \BibitemOpen
  \bibfield  {author} {\bibinfo {author} {\bibfnamefont {A.}~\bibnamefont
  {{Harada}}}\ and\ \bibinfo {author} {\bibfnamefont {H.}~\bibnamefont
  {{Nagakura}}},\ }\href {\doibase 10.3847/1538-4357/ac38a0} {\bibfield
  {journal} {\bibinfo  {journal} {Astrophys. J.}\ }\textbf {\bibinfo {volume}
  {924}},\ \bibinfo {eid} {109} (\bibinfo {year} {2022})},\ \Eprint
  {http://arxiv.org/abs/2110.08291} {arXiv:2110.08291 [astro-ph.HE]}
  \BibitemShut {NoStop}%
\bibitem [{\citenamefont {Richers}(2022)}]{richers2022evaluating}%
  \BibitemOpen
  \bibfield  {author} {\bibinfo {author} {\bibfnamefont {S.}~\bibnamefont
  {Richers}},\ }\href {\doibase 10.1103/PhysRevD.106.083005} {\bibfield
  {journal} {\bibinfo  {journal} {Phys. Rev. D}\ }\textbf {\bibinfo {volume}
  {106}},\ \bibinfo {pages} {083005} (\bibinfo {year} {2022})},\ \Eprint
  {http://arxiv.org/abs/2206.08444} {arXiv:2206.08444} \BibitemShut {NoStop}%
\bibitem [{\citenamefont {{Martin}}\ \emph {et~al.}(2020)\citenamefont
  {{Martin}}, \citenamefont {{Yi}},\ and\ \citenamefont
  {{Duan}}}]{martin2020dynamic}%
  \BibitemOpen
  \bibfield  {author} {\bibinfo {author} {\bibfnamefont {J.~D.}\ \bibnamefont
  {{Martin}}}, \bibinfo {author} {\bibfnamefont {C.}~\bibnamefont {{Yi}}}, \
  and\ \bibinfo {author} {\bibfnamefont {H.}~\bibnamefont {{Duan}}},\ }\href
  {\doibase 10.1016/j.physletb.2019.135088} {\bibfield  {journal} {\bibinfo
  {journal} {Phys. Lett. B}\ }\textbf {\bibinfo {volume} {800}},\ \bibinfo
  {eid} {135088} (\bibinfo {year} {2020})},\ \Eprint
  {http://arxiv.org/abs/1909.05225} {arXiv:1909.05225 [hep-ph]} \BibitemShut
  {NoStop}%
\bibitem [{\citenamefont {{Bhattacharyya}}\ and\ \citenamefont
  {{Dasgupta}}(2021)}]{bhattacharyya2021fast}%
  \BibitemOpen
  \bibfield  {author} {\bibinfo {author} {\bibfnamefont {S.}~\bibnamefont
  {{Bhattacharyya}}}\ and\ \bibinfo {author} {\bibfnamefont {B.}~\bibnamefont
  {{Dasgupta}}},\ }\href {\doibase 10.1103/PhysRevLett.126.061302} {\bibfield
  {journal} {\bibinfo  {journal} {Phys. Rev. Lett.}\ }\textbf {\bibinfo
  {volume} {126}},\ \bibinfo {eid} {061302} (\bibinfo {year} {2021})},\ \Eprint
  {http://arxiv.org/abs/2009.03337} {arXiv:2009.03337 [hep-ph]} \BibitemShut
  {NoStop}%
\bibitem [{\citenamefont {Bhattacharyya}\ and\ \citenamefont
  {Dasgupta}(2020)}]{bhattacharyya2020late}%
  \BibitemOpen
  \bibfield  {author} {\bibinfo {author} {\bibfnamefont {S.}~\bibnamefont
  {Bhattacharyya}}\ and\ \bibinfo {author} {\bibfnamefont {B.}~\bibnamefont
  {Dasgupta}},\ }\href {\doibase 10.1103/PhysRevD.102.063018} {\bibfield
  {journal} {\bibinfo  {journal} {Phys. Rev. D}\ }\textbf {\bibinfo {volume}
  {102}},\ \bibinfo {pages} {063018} (\bibinfo {year} {2020})},\ \Eprint
  {http://arxiv.org/abs/2005.00459} {arXiv:2005.00459} \BibitemShut {NoStop}%
\bibitem [{\citenamefont {{Wu}}\ \emph {et~al.}(2021)\citenamefont {{Wu}},
  \citenamefont {{George}}, \citenamefont {{Lin}},\ and\ \citenamefont
  {{Xiong}}}]{wu2021collective}%
  \BibitemOpen
  \bibfield  {author} {\bibinfo {author} {\bibfnamefont {M.-R.}\ \bibnamefont
  {{Wu}}}, \bibinfo {author} {\bibfnamefont {M.}~\bibnamefont {{George}}},
  \bibinfo {author} {\bibfnamefont {C.-Y.}\ \bibnamefont {{Lin}}}, \ and\
  \bibinfo {author} {\bibfnamefont {Z.}~\bibnamefont {{Xiong}}},\ }\href
  {\doibase 10.1103/PhysRevD.104.103003} {\bibfield  {journal} {\bibinfo
  {journal} {Phys. Rev. D}\ }\textbf {\bibinfo {volume} {104}},\ \bibinfo {eid}
  {103003} (\bibinfo {year} {2021})},\ \Eprint
  {http://arxiv.org/abs/2108.09886} {arXiv:2108.09886 [hep-ph]} \BibitemShut
  {NoStop}%
\bibitem [{\citenamefont {Richers}\ \emph
  {et~al.}(2021{\natexlab{a}})\citenamefont {Richers}, \citenamefont {Willcox},
  \citenamefont {Ford},\ and\ \citenamefont {Myers}}]{richers2021particle}%
  \BibitemOpen
  \bibfield  {author} {\bibinfo {author} {\bibfnamefont {S.}~\bibnamefont
  {Richers}}, \bibinfo {author} {\bibfnamefont {D.~E.}\ \bibnamefont
  {Willcox}}, \bibinfo {author} {\bibfnamefont {N.~M.}\ \bibnamefont {Ford}}, \
  and\ \bibinfo {author} {\bibfnamefont {A.}~\bibnamefont {Myers}},\ }\href
  {\doibase 10.1103/PhysRevD.103.083013} {\bibfield  {journal} {\bibinfo
  {journal} {Phys. Rev. D}\ }\textbf {\bibinfo {volume} {103}},\ \bibinfo
  {pages} {083013} (\bibinfo {year} {2021}{\natexlab{a}})},\ \Eprint
  {http://arxiv.org/abs/2101.02745} {arXiv:2101.02745} \BibitemShut {NoStop}%
\bibitem [{\citenamefont {Richers}\ \emph
  {et~al.}(2021{\natexlab{b}})\citenamefont {Richers}, \citenamefont
  {Willcox},\ and\ \citenamefont {Ford}}]{richers2021neutrino}%
  \BibitemOpen
  \bibfield  {author} {\bibinfo {author} {\bibfnamefont {S.}~\bibnamefont
  {Richers}}, \bibinfo {author} {\bibfnamefont {D.}~\bibnamefont {Willcox}}, \
  and\ \bibinfo {author} {\bibfnamefont {N.}~\bibnamefont {Ford}},\ }\href
  {\doibase 10.1103/PhysRevD.104.103023} {\bibfield  {journal} {\bibinfo
  {journal} {Phys. Rev. D}\ }\textbf {\bibinfo {volume} {104}},\ \bibinfo
  {pages} {103023} (\bibinfo {year} {2021}{\natexlab{b}})},\ \Eprint
  {http://arxiv.org/abs/2109.08631} {arXiv:2109.08631} \BibitemShut {NoStop}%
\bibitem [{\citenamefont {{Zaizen}}\ and\ \citenamefont
  {{Morinaga}}(2021)}]{zaizen2021nonlinear}%
  \BibitemOpen
  \bibfield  {author} {\bibinfo {author} {\bibfnamefont {M.}~\bibnamefont
  {{Zaizen}}}\ and\ \bibinfo {author} {\bibfnamefont {T.}~\bibnamefont
  {{Morinaga}}},\ }\href {\doibase 10.1103/PhysRevD.104.083035} {\bibfield
  {journal} {\bibinfo  {journal} {Phys. Rev. D}\ }\textbf {\bibinfo {volume}
  {104}},\ \bibinfo {eid} {083035} (\bibinfo {year} {2021})},\ \Eprint
  {http://arxiv.org/abs/2104.10532} {arXiv:2104.10532 [hep-ph]} \BibitemShut
  {NoStop}%
\bibitem [{\citenamefont {{Abbar}}\ and\ \citenamefont
  {{Capozzi}}(2022)}]{abbar2022suppression}%
  \BibitemOpen
  \bibfield  {author} {\bibinfo {author} {\bibfnamefont {S.}~\bibnamefont
  {{Abbar}}}\ and\ \bibinfo {author} {\bibfnamefont {F.}~\bibnamefont
  {{Capozzi}}},\ }\href {\doibase 10.1088/1475-7516/2022/03/051} {\bibfield
  {journal} {\bibinfo  {journal} {J. Cosmol. Astropart. Phys.}\ }\textbf
  {\bibinfo {volume} {2022}},\ \bibinfo {eid} {051} (\bibinfo {year} {2022})},\
  \Eprint {http://arxiv.org/abs/2111.14880} {arXiv:2111.14880 [astro-ph.HE]}
  \BibitemShut {NoStop}%
\bibitem [{\citenamefont {Richers}\ \emph {et~al.}(2022)\citenamefont
  {Richers}, \citenamefont {Duan}, \citenamefont {Wu}, \citenamefont
  {Bhattacharyya}, \citenamefont {Zaizen}, \citenamefont {George},
  \citenamefont {Lin},\ and\ \citenamefont {Xiong}}]{richers2022code}%
  \BibitemOpen
  \bibfield  {author} {\bibinfo {author} {\bibfnamefont {S.}~\bibnamefont
  {Richers}}, \bibinfo {author} {\bibfnamefont {H.}~\bibnamefont {Duan}},
  \bibinfo {author} {\bibfnamefont {M.-R.}\ \bibnamefont {Wu}}, \bibinfo
  {author} {\bibfnamefont {S.}~\bibnamefont {Bhattacharyya}}, \bibinfo {author}
  {\bibfnamefont {M.}~\bibnamefont {Zaizen}}, \bibinfo {author} {\bibfnamefont
  {M.}~\bibnamefont {George}}, \bibinfo {author} {\bibfnamefont {C.-Y.}\
  \bibnamefont {Lin}}, \ and\ \bibinfo {author} {\bibfnamefont
  {Z.}~\bibnamefont {Xiong}},\ }\href {\doibase 10.1103/PhysRevD.106.043011}
  {\bibfield  {journal} {\bibinfo  {journal} {Phys. Rev. D}\ }\textbf {\bibinfo
  {volume} {106}},\ \bibinfo {pages} {043011} (\bibinfo {year} {2022})},\
  \Eprint {http://arxiv.org/abs/2205.06282} {arXiv:2205.06282} \BibitemShut
  {NoStop}%
\bibitem [{\citenamefont {Bhattacharyya}\ and\ \citenamefont
  {Dasgupta}(2022)}]{bhattacharyya2022elaborating}%
  \BibitemOpen
  \bibfield  {author} {\bibinfo {author} {\bibfnamefont {S.}~\bibnamefont
  {Bhattacharyya}}\ and\ \bibinfo {author} {\bibfnamefont {B.}~\bibnamefont
  {Dasgupta}},\ }\href {\doibase 10.1103/PhysRevD.106.103039} {\bibfield
  {journal} {\bibinfo  {journal} {Phys. Rev. D}\ }\textbf {\bibinfo {volume}
  {106}},\ \bibinfo {pages} {103039} (\bibinfo {year} {2022})},\ \Eprint
  {http://arxiv.org/abs/2205.05129} {arXiv:2205.05129} \BibitemShut {NoStop}%
\bibitem [{\citenamefont {Grohs}\ \emph {et~al.}(2023)\citenamefont {Grohs},
  \citenamefont {Richers}, \citenamefont {Couch}, \citenamefont {Foucart},
  \citenamefont {Kneller},\ and\ \citenamefont
  {McLaughlin}}]{grohs2023neutrino}%
  \BibitemOpen
  \bibfield  {author} {\bibinfo {author} {\bibfnamefont {E.}~\bibnamefont
  {Grohs}}, \bibinfo {author} {\bibfnamefont {S.}~\bibnamefont {Richers}},
  \bibinfo {author} {\bibfnamefont {S.~M.}\ \bibnamefont {Couch}}, \bibinfo
  {author} {\bibfnamefont {F.}~\bibnamefont {Foucart}}, \bibinfo {author}
  {\bibfnamefont {J.~P.}\ \bibnamefont {Kneller}}, \ and\ \bibinfo {author}
  {\bibfnamefont {G.~C.}\ \bibnamefont {McLaughlin}},\ }\href {\doibase
  10.1016/j.physletb.2023.138210} {\bibfield  {journal} {\bibinfo  {journal}
  {Phys. Lett. B}\ }\textbf {\bibinfo {volume} {846}},\ \bibinfo {pages}
  {138210} (\bibinfo {year} {2023})},\ \Eprint
  {http://arxiv.org/abs/2207.02214} {arXiv:2207.02214} \BibitemShut {NoStop}%
\bibitem [{\citenamefont {{Zaizen}}\ and\ \citenamefont
  {{Nagakura}}(2023{\natexlab{a}})}]{zaizen2023simple}%
  \BibitemOpen
  \bibfield  {author} {\bibinfo {author} {\bibfnamefont {M.}~\bibnamefont
  {{Zaizen}}}\ and\ \bibinfo {author} {\bibfnamefont {H.}~\bibnamefont
  {{Nagakura}}},\ }\href {\doibase 10.1103/PhysRevD.107.103022} {\bibfield
  {journal} {\bibinfo  {journal} {Phys. Rev. D}\ }\textbf {\bibinfo {volume}
  {107}},\ \bibinfo {eid} {103022} (\bibinfo {year} {2023}{\natexlab{a}})},\
  \Eprint {http://arxiv.org/abs/2211.09343} {arXiv:2211.09343 [astro-ph.HE]}
  \BibitemShut {NoStop}%
\bibitem [{\citenamefont {{Zaizen}}\ and\ \citenamefont
  {{Nagakura}}(2023{\natexlab{b}})}]{zaizen2023characterizing}%
  \BibitemOpen
  \bibfield  {author} {\bibinfo {author} {\bibfnamefont {M.}~\bibnamefont
  {{Zaizen}}}\ and\ \bibinfo {author} {\bibfnamefont {H.}~\bibnamefont
  {{Nagakura}}},\ }\href {\doibase 10.1103/PhysRevD.107.123021} {\bibfield
  {journal} {\bibinfo  {journal} {Phys. Rev. D}\ }\textbf {\bibinfo {volume}
  {107}},\ \bibinfo {eid} {123021} (\bibinfo {year} {2023}{\natexlab{b}})},\
  \Eprint {http://arxiv.org/abs/2304.05044} {arXiv:2304.05044 [astro-ph.HE]}
  \BibitemShut {NoStop}%
\bibitem [{\citenamefont {Xiong}\ \emph {et~al.}(2023)\citenamefont {Xiong},
  \citenamefont {Wu}, \citenamefont {Abbar}, \citenamefont {Bhattacharyya},
  \citenamefont {George},\ and\ \citenamefont {Lin}}]{xiong2023evaluating}%
  \BibitemOpen
  \bibfield  {author} {\bibinfo {author} {\bibfnamefont {Z.}~\bibnamefont
  {Xiong}}, \bibinfo {author} {\bibfnamefont {M.-R.}\ \bibnamefont {Wu}},
  \bibinfo {author} {\bibfnamefont {S.}~\bibnamefont {Abbar}}, \bibinfo
  {author} {\bibfnamefont {S.}~\bibnamefont {Bhattacharyya}}, \bibinfo {author}
  {\bibfnamefont {M.}~\bibnamefont {George}}, \ and\ \bibinfo {author}
  {\bibfnamefont {C.-Y.}\ \bibnamefont {Lin}},\ }\href {\doibase
  10.1103/PhysRevD.108.063003} {\bibfield  {journal} {\bibinfo  {journal}
  {Phys. Rev. D}\ }\textbf {\bibinfo {volume} {108}},\ \bibinfo {pages}
  {063003} (\bibinfo {year} {2023})}\BibitemShut {NoStop}%
\bibitem [{\citenamefont {Nagakura}\ \emph {et~al.}(2024)\citenamefont
  {Nagakura}, \citenamefont {Johns},\ and\ \citenamefont
  {Zaizen}}]{nagakura2023bgk}%
  \BibitemOpen
  \bibfield  {author} {\bibinfo {author} {\bibfnamefont {H.}~\bibnamefont
  {Nagakura}}, \bibinfo {author} {\bibfnamefont {L.}~\bibnamefont {Johns}}, \
  and\ \bibinfo {author} {\bibfnamefont {M.}~\bibnamefont {Zaizen}},\ }\href
  {\doibase 10.1103/PhysRevD.109.083013} {\bibfield  {journal} {\bibinfo
  {journal} {Phys. Rev. D}\ }\textbf {\bibinfo {volume} {109}},\ \bibinfo
  {pages} {083013} (\bibinfo {year} {2024})},\ \Eprint
  {http://arxiv.org/abs/2312.16285} {arXiv:2312.16285 [astro-ph.HE]}
  \BibitemShut {NoStop}%
\bibitem [{\citenamefont {Grohs}\ \emph {et~al.}(2024)\citenamefont {Grohs},
  \citenamefont {Richers}, \citenamefont {Couch}, \citenamefont {Foucart},
  \citenamefont {Froustey}, \citenamefont {Kneller},\ and\ \citenamefont
  {McLaughlin}}]{grohs2024two}%
  \BibitemOpen
  \bibfield  {author} {\bibinfo {author} {\bibfnamefont {E.}~\bibnamefont
  {Grohs}}, \bibinfo {author} {\bibfnamefont {S.}~\bibnamefont {Richers}},
  \bibinfo {author} {\bibfnamefont {S.~M.}\ \bibnamefont {Couch}}, \bibinfo
  {author} {\bibfnamefont {F.}~\bibnamefont {Foucart}}, \bibinfo {author}
  {\bibfnamefont {J.}~\bibnamefont {Froustey}}, \bibinfo {author}
  {\bibfnamefont {J.~P.}\ \bibnamefont {Kneller}}, \ and\ \bibinfo {author}
  {\bibfnamefont {G.~C.}\ \bibnamefont {McLaughlin}},\ }\href {\doibase
  10.3847/1538-4357/ad13f2} {\bibfield  {journal} {\bibinfo  {journal}
  {Astrophys. J.}\ }\textbf {\bibinfo {volume} {963}},\ \bibinfo {pages} {11}
  (\bibinfo {year} {2024})},\ \Eprint {http://arxiv.org/abs/2309.00972}
  {arXiv:2309.00972 [astro-ph.HE]} \BibitemShut {NoStop}%
\bibitem [{\citenamefont {Froustey}\ \emph {et~al.}(2024)\citenamefont
  {Froustey}, \citenamefont {Richers}, \citenamefont {Grohs}, \citenamefont
  {Flynn}, \citenamefont {Foucart}, \citenamefont {Kneller},\ and\
  \citenamefont {McLaughlin}}]{froustey2024neutrino}%
  \BibitemOpen
  \bibfield  {author} {\bibinfo {author} {\bibfnamefont {J.}~\bibnamefont
  {Froustey}}, \bibinfo {author} {\bibfnamefont {S.}~\bibnamefont {Richers}},
  \bibinfo {author} {\bibfnamefont {E.}~\bibnamefont {Grohs}}, \bibinfo
  {author} {\bibfnamefont {S.~D.}\ \bibnamefont {Flynn}}, \bibinfo {author}
  {\bibfnamefont {F.}~\bibnamefont {Foucart}}, \bibinfo {author} {\bibfnamefont
  {J.~P.}\ \bibnamefont {Kneller}}, \ and\ \bibinfo {author} {\bibfnamefont
  {G.~C.}\ \bibnamefont {McLaughlin}},\ }\href {\doibase
  10.1103/PhysRevD.109.043046} {\bibfield  {journal} {\bibinfo  {journal}
  {Phys. Rev. D}\ }\textbf {\bibinfo {volume} {109}},\ \bibinfo {pages}
  {043046} (\bibinfo {year} {2024})},\ \Eprint
  {http://arxiv.org/abs/2311.11968} {arXiv:2311.11968 [astro-ph.HE]}
  \BibitemShut {NoStop}%
\bibitem [{\citenamefont {{Xiong}}\ and\ \citenamefont
  {{Qian}}(2021)}]{xiong2021stationary}%
  \BibitemOpen
  \bibfield  {author} {\bibinfo {author} {\bibfnamefont {Z.}~\bibnamefont
  {{Xiong}}}\ and\ \bibinfo {author} {\bibfnamefont {Y.-Z.}\ \bibnamefont
  {{Qian}}},\ }\href {\doibase 10.1016/j.physletb.2021.136550} {\bibfield
  {journal} {\bibinfo  {journal} {Phys. Lett. B}\ }\textbf {\bibinfo {volume}
  {820}},\ \bibinfo {eid} {136550} (\bibinfo {year} {2021})},\ \Eprint
  {http://arxiv.org/abs/2104.05618} {arXiv:2104.05618 [astro-ph.HE]}
  \BibitemShut {NoStop}%
\bibitem [{\citenamefont {{Abbar}}\ \emph {et~al.}(2024)\citenamefont
  {{Abbar}}, \citenamefont {{Wu}},\ and\ \citenamefont
  {{Xiong}}}]{abbar2024physics}%
  \BibitemOpen
  \bibfield  {author} {\bibinfo {author} {\bibfnamefont {S.}~\bibnamefont
  {{Abbar}}}, \bibinfo {author} {\bibfnamefont {M.-R.}\ \bibnamefont {{Wu}}}, \
  and\ \bibinfo {author} {\bibfnamefont {Z.}~\bibnamefont {{Xiong}}},\ }\href
  {\doibase 10.1103/PhysRevD.109.043024} {\bibfield  {journal} {\bibinfo
  {journal} {Phys. Rev. D}\ }\textbf {\bibinfo {volume} {109}},\ \bibinfo {eid}
  {043024} (\bibinfo {year} {2024})},\ \Eprint
  {http://arxiv.org/abs/2311.15656} {arXiv:2311.15656} \BibitemShut {NoStop}%
\bibitem [{\citenamefont {Abbar}\ \emph {et~al.}(2024)\citenamefont {Abbar},
  \citenamefont {Wu},\ and\ \citenamefont {Xiong}}]{abbar2024applications}%
  \BibitemOpen
  \bibfield  {author} {\bibinfo {author} {\bibfnamefont {S.}~\bibnamefont
  {Abbar}}, \bibinfo {author} {\bibfnamefont {M.-R.}\ \bibnamefont {Wu}}, \
  and\ \bibinfo {author} {\bibfnamefont {Z.}~\bibnamefont {Xiong}},\ }\href
  {\doibase 10.1103/PhysRevD.109.083019} {\bibfield  {journal} {\bibinfo
  {journal} {Phys. Rev. D}\ }\textbf {\bibinfo {volume} {109}},\ \bibinfo
  {pages} {083019} (\bibinfo {year} {2024})},\ \Eprint
  {http://arxiv.org/abs/2401.17424} {arXiv:2401.17424 [astro-ph.HE]}
  \BibitemShut {NoStop}%
\bibitem [{\citenamefont {{Capozzi}}\ \emph {et~al.}(2019)\citenamefont
  {{Capozzi}}, \citenamefont {{Dasgupta}}, \citenamefont {{Mirizzi}},
  \citenamefont {{Sen}},\ and\ \citenamefont
  {{Sigl}}}]{capozzi2019collisional}%
  \BibitemOpen
  \bibfield  {author} {\bibinfo {author} {\bibfnamefont {F.}~\bibnamefont
  {{Capozzi}}}, \bibinfo {author} {\bibfnamefont {B.}~\bibnamefont
  {{Dasgupta}}}, \bibinfo {author} {\bibfnamefont {A.}~\bibnamefont
  {{Mirizzi}}}, \bibinfo {author} {\bibfnamefont {M.}~\bibnamefont {{Sen}}}, \
  and\ \bibinfo {author} {\bibfnamefont {G.}~\bibnamefont {{Sigl}}},\ }\href
  {\doibase 10.1103/PhysRevLett.122.091101} {\bibfield  {journal} {\bibinfo
  {journal} {Phys. Rev. Lett.}\ }\textbf {\bibinfo {volume} {122}},\ \bibinfo
  {eid} {091101} (\bibinfo {year} {2019})},\ \Eprint
  {http://arxiv.org/abs/1808.06618} {arXiv:1808.06618 [hep-ph]} \BibitemShut
  {NoStop}%
\bibitem [{\citenamefont {Nagakura}\ and\ \citenamefont
  {Zaizen}(2022)}]{nagakura2022time}%
  \BibitemOpen
  \bibfield  {author} {\bibinfo {author} {\bibfnamefont {H.}~\bibnamefont
  {Nagakura}}\ and\ \bibinfo {author} {\bibfnamefont {M.}~\bibnamefont
  {Zaizen}},\ }\href {\doibase 10.1103/PhysRevLett.129.261101} {\bibfield
  {journal} {\bibinfo  {journal} {Phys. Rev. Lett.}\ }\textbf {\bibinfo
  {volume} {129}},\ \bibinfo {pages} {261101} (\bibinfo {year} {2022})},\
  \Eprint {http://arxiv.org/abs/2206.04097} {arXiv:2206.04097 [astro-ph.HE]}
  \BibitemShut {NoStop}%
\bibitem [{\citenamefont {{Xiong}}\ \emph {et~al.}(2023)\citenamefont
  {{Xiong}}, \citenamefont {{Wu}}, \citenamefont {{Mart{\'\i}nez-Pinedo}},
  \citenamefont {{Fischer}}, \citenamefont {{George}}, \citenamefont {{Lin}},\
  and\ \citenamefont {{Johns}}}]{xiong2023evolution}%
  \BibitemOpen
  \bibfield  {author} {\bibinfo {author} {\bibfnamefont {Z.}~\bibnamefont
  {{Xiong}}}, \bibinfo {author} {\bibfnamefont {M.-R.}\ \bibnamefont {{Wu}}},
  \bibinfo {author} {\bibfnamefont {G.}~\bibnamefont {{Mart{\'\i}nez-Pinedo}}},
  \bibinfo {author} {\bibfnamefont {T.}~\bibnamefont {{Fischer}}}, \bibinfo
  {author} {\bibfnamefont {M.}~\bibnamefont {{George}}}, \bibinfo {author}
  {\bibfnamefont {C.-Y.}\ \bibnamefont {{Lin}}}, \ and\ \bibinfo {author}
  {\bibfnamefont {L.}~\bibnamefont {{Johns}}},\ }\href {\doibase
  10.1103/PhysRevD.107.083016} {\bibfield  {journal} {\bibinfo  {journal}
  {Phys. Rev. D}\ }\textbf {\bibinfo {volume} {107}},\ \bibinfo {eid} {083016}
  (\bibinfo {year} {2023})},\ \Eprint {http://arxiv.org/abs/2210.08254}
  {arXiv:2210.08254 [astro-ph.HE]} \BibitemShut {NoStop}%
\bibitem [{\citenamefont {Nagakura}\ and\ \citenamefont
  {Zaizen}(2023)}]{nagakura2023basic}%
  \BibitemOpen
  \bibfield  {author} {\bibinfo {author} {\bibfnamefont {H.}~\bibnamefont
  {Nagakura}}\ and\ \bibinfo {author} {\bibfnamefont {M.}~\bibnamefont
  {Zaizen}},\ }\href {\doibase 10.1103/PhysRevD.108.123003} {\bibfield
  {journal} {\bibinfo  {journal} {Phys. Rev. D}\ }\textbf {\bibinfo {volume}
  {108}},\ \bibinfo {pages} {123003} (\bibinfo {year} {2023})},\ \Eprint
  {http://arxiv.org/abs/2308.14800} {arXiv:2308.14800 [astro-ph.HE]}
  \BibitemShut {NoStop}%
\bibitem [{\citenamefont {{Nagakura}}(2023)}]{nagakura2023roles}%
  \BibitemOpen
  \bibfield  {author} {\bibinfo {author} {\bibfnamefont {H.}~\bibnamefont
  {{Nagakura}}},\ }\href {\doibase 10.1103/PhysRevLett.130.211401} {\bibfield
  {journal} {\bibinfo  {journal} {Phys. Rev. Lett.}\ }\textbf {\bibinfo
  {volume} {130}},\ \bibinfo {eid} {211401} (\bibinfo {year} {2023})},\ \Eprint
  {http://arxiv.org/abs/2301.10785} {arXiv:2301.10785 [astro-ph.HE]}
  \BibitemShut {NoStop}%
\bibitem [{\citenamefont {Nagakura}(2023)}]{nagakura2023global}%
  \BibitemOpen
  \bibfield  {author} {\bibinfo {author} {\bibfnamefont {H.}~\bibnamefont
  {Nagakura}},\ }\href {\doibase 10.1103/PhysRevD.108.103014} {\bibfield
  {journal} {\bibinfo  {journal} {Phys. Rev. D}\ }\textbf {\bibinfo {volume}
  {108}},\ \bibinfo {pages} {103014} (\bibinfo {year} {2023})},\ \Eprint
  {http://arxiv.org/abs/2306.10108} {arXiv:2306.10108 [astro-ph.HE]}
  \BibitemShut {NoStop}%
\bibitem [{\citenamefont {{Xiong}}\ \emph {et~al.}(2024)\citenamefont
  {{Xiong}}, \citenamefont {{Wu}}, \citenamefont {{George}}, \citenamefont
  {{Lin}}, \citenamefont {{Khosravi Largani}}, \citenamefont {{Fischer}},\ and\
  \citenamefont {{Mart{\'\i}nez-Pinedo}}}]{xiong2024fast}%
  \BibitemOpen
  \bibfield  {author} {\bibinfo {author} {\bibfnamefont {Z.}~\bibnamefont
  {{Xiong}}}, \bibinfo {author} {\bibfnamefont {M.-R.}\ \bibnamefont {{Wu}}},
  \bibinfo {author} {\bibfnamefont {M.}~\bibnamefont {{George}}}, \bibinfo
  {author} {\bibfnamefont {C.-Y.}\ \bibnamefont {{Lin}}}, \bibinfo {author}
  {\bibfnamefont {N.}~\bibnamefont {{Khosravi Largani}}}, \bibinfo {author}
  {\bibfnamefont {T.}~\bibnamefont {{Fischer}}}, \ and\ \bibinfo {author}
  {\bibfnamefont {G.}~\bibnamefont {{Mart{\'\i}nez-Pinedo}}},\ }\href {\doibase
  10.1103/PhysRevD.109.123008} {\bibfield  {journal} {\bibinfo  {journal}
  {Phys. Rev. D}\ }\textbf {\bibinfo {volume} {109}},\ \bibinfo {pages}
  {123008} (\bibinfo {year} {2024})},\ \Eprint
  {http://arxiv.org/abs/2402.19252} {2402.19252} \BibitemShut {NoStop}%
\bibitem [{\citenamefont {Shalgar}\ and\ \citenamefont
  {Tamborra}(2023)}]{shalgar2023supernova}%
  \BibitemOpen
  \bibfield  {author} {\bibinfo {author} {\bibfnamefont {S.}~\bibnamefont
  {Shalgar}}\ and\ \bibinfo {author} {\bibfnamefont {I.}~\bibnamefont
  {Tamborra}},\ }\href {\doibase 10.1103/PhysRevD.108.043006} {\bibfield
  {journal} {\bibinfo  {journal} {Phys. Rev. D}\ }\textbf {\bibinfo {volume}
  {108}},\ \bibinfo {pages} {043006} (\bibinfo {year} {2023})},\ \Eprint
  {http://arxiv.org/abs/2206.00676} {arXiv:2206.00676 [astro-ph.HE]}
  \BibitemShut {NoStop}%
\bibitem [{\citenamefont {{Shalgar}}\ and\ \citenamefont
  {{Tamborra}}(2023)}]{shalgar2023neutrino}%
  \BibitemOpen
  \bibfield  {author} {\bibinfo {author} {\bibfnamefont {S.}~\bibnamefont
  {{Shalgar}}}\ and\ \bibinfo {author} {\bibfnamefont {I.}~\bibnamefont
  {{Tamborra}}},\ }\href {\doibase 10.1103/PhysRevD.107.063025} {\bibfield
  {journal} {\bibinfo  {journal} {Phys. Rev. D}\ }\textbf {\bibinfo {volume}
  {107}},\ \bibinfo {eid} {063025} (\bibinfo {year} {2023})},\ \Eprint
  {http://arxiv.org/abs/2207.04058} {arXiv:2207.04058 [astro-ph.HE]}
  \BibitemShut {NoStop}%
\bibitem [{\citenamefont {Cornelius}\ \emph {et~al.}(2024)\citenamefont
  {Cornelius}, \citenamefont {Shalgar},\ and\ \citenamefont
  {Tamborra}}]{cornelius2024perturbing}%
  \BibitemOpen
  \bibfield  {author} {\bibinfo {author} {\bibfnamefont {M.}~\bibnamefont
  {Cornelius}}, \bibinfo {author} {\bibfnamefont {S.}~\bibnamefont {Shalgar}},
  \ and\ \bibinfo {author} {\bibfnamefont {I.}~\bibnamefont {Tamborra}},\
  }\href {\doibase 10.1088/1475-7516/2024/02/038} {\bibfield  {journal}
  {\bibinfo  {journal} {JCAP}\ }\textbf {\bibinfo {volume} {02}},\ \bibinfo
  {pages} {038} (\bibinfo {year} {2024})},\ \Eprint
  {http://arxiv.org/abs/2312.03839} {arXiv:2312.03839 [astro-ph.HE]}
  \BibitemShut {NoStop}%
\bibitem [{\citenamefont {{Li}}\ and\ \citenamefont
  {{Siegel}}(2021)}]{li2021neutrino}%
  \BibitemOpen
  \bibfield  {author} {\bibinfo {author} {\bibfnamefont {X.}~\bibnamefont
  {{Li}}}\ and\ \bibinfo {author} {\bibfnamefont {D.~M.}\ \bibnamefont
  {{Siegel}}},\ }\href {\doibase 10.1103/PhysRevLett.126.251101} {\bibfield
  {journal} {\bibinfo  {journal} {Phys. Rev. Lett.}\ }\textbf {\bibinfo
  {volume} {126}},\ \bibinfo {eid} {251101} (\bibinfo {year} {2021})},\ \Eprint
  {http://arxiv.org/abs/2103.02616} {arXiv:2103.02616 [astro-ph.HE]}
  \BibitemShut {NoStop}%
\bibitem [{\citenamefont {Just}\ \emph {et~al.}(2022)\citenamefont {Just},
  \citenamefont {Abbar}, \citenamefont {Wu}, \citenamefont {Tamborra},
  \citenamefont {Janka},\ and\ \citenamefont {Capozzi}}]{just2022fast}%
  \BibitemOpen
  \bibfield  {author} {\bibinfo {author} {\bibfnamefont {O.}~\bibnamefont
  {Just}}, \bibinfo {author} {\bibfnamefont {S.}~\bibnamefont {Abbar}},
  \bibinfo {author} {\bibfnamefont {M.~R.}\ \bibnamefont {Wu}}, \bibinfo
  {author} {\bibfnamefont {I.}~\bibnamefont {Tamborra}}, \bibinfo {author}
  {\bibfnamefont {H.~T.}\ \bibnamefont {Janka}}, \ and\ \bibinfo {author}
  {\bibfnamefont {F.}~\bibnamefont {Capozzi}},\ }\href {\doibase
  10.1103/PhysRevD.105.083024} {\bibfield  {journal} {\bibinfo  {journal}
  {Phys. Rev. D}\ }\textbf {\bibinfo {volume} {105}},\ \bibinfo {pages}
  {083024} (\bibinfo {year} {2022})},\ \Eprint
  {http://arxiv.org/abs/2203.16559} {arXiv:2203.16559} \BibitemShut {NoStop}%
\bibitem [{\citenamefont {{Fern{\'a}ndez}}\ \emph {et~al.}(2022)\citenamefont
  {{Fern{\'a}ndez}}, \citenamefont {{Richers}}, \citenamefont {{Mulyk}},\ and\
  \citenamefont {{Fahlman}}}]{fernandez2023fast}%
  \BibitemOpen
  \bibfield  {author} {\bibinfo {author} {\bibfnamefont {R.}~\bibnamefont
  {{Fern{\'a}ndez}}}, \bibinfo {author} {\bibfnamefont {S.}~\bibnamefont
  {{Richers}}}, \bibinfo {author} {\bibfnamefont {N.}~\bibnamefont {{Mulyk}}},
  \ and\ \bibinfo {author} {\bibfnamefont {S.}~\bibnamefont {{Fahlman}}},\
  }\href {\doibase 10.1103/PhysRevD.106.103003} {\bibfield  {journal} {\bibinfo
   {journal} {Phys. Rev. D}\ }\textbf {\bibinfo {volume} {106}},\ \bibinfo
  {eid} {103003} (\bibinfo {year} {2022})},\ \Eprint
  {http://arxiv.org/abs/2207.10680} {arXiv:2207.10680 [astro-ph.HE]}
  \BibitemShut {NoStop}%
\bibitem [{\citenamefont {Ehring}\ \emph
  {et~al.}(2023{\natexlab{a}})\citenamefont {Ehring}, \citenamefont {Abbar},
  \citenamefont {Janka}, \citenamefont {Raffelt},\ and\ \citenamefont
  {Tamborra}}]{ehring2023fast}%
  \BibitemOpen
  \bibfield  {author} {\bibinfo {author} {\bibfnamefont {J.}~\bibnamefont
  {Ehring}}, \bibinfo {author} {\bibfnamefont {S.}~\bibnamefont {Abbar}},
  \bibinfo {author} {\bibfnamefont {H.~T.}\ \bibnamefont {Janka}}, \bibinfo
  {author} {\bibfnamefont {G.}~\bibnamefont {Raffelt}}, \ and\ \bibinfo
  {author} {\bibfnamefont {I.}~\bibnamefont {Tamborra}},\ }\href {\doibase
  10.1103/PhysRevD.107.103034} {\bibfield  {journal} {\bibinfo  {journal}
  {Phys. Rev. D}\ }\textbf {\bibinfo {volume} {107}},\ \bibinfo {pages}
  {103034} (\bibinfo {year} {2023}{\natexlab{a}})},\ \Eprint
  {http://arxiv.org/abs/2301.11938} {arXiv:2301.11938} \BibitemShut {NoStop}%
\bibitem [{\citenamefont {Ehring}\ \emph
  {et~al.}(2023{\natexlab{b}})\citenamefont {Ehring}, \citenamefont {Abbar},
  \citenamefont {Janka}, \citenamefont {Raffelt},\ and\ \citenamefont
  {Tamborra}}]{ehring2023fast2}%
  \BibitemOpen
  \bibfield  {author} {\bibinfo {author} {\bibfnamefont {J.}~\bibnamefont
  {Ehring}}, \bibinfo {author} {\bibfnamefont {S.}~\bibnamefont {Abbar}},
  \bibinfo {author} {\bibfnamefont {H.~T.}\ \bibnamefont {Janka}}, \bibinfo
  {author} {\bibfnamefont {G.}~\bibnamefont {Raffelt}}, \ and\ \bibinfo
  {author} {\bibfnamefont {I.}~\bibnamefont {Tamborra}},\ }\href {\doibase
  10.1103/PhysRevLett.131.061401} {\bibfield  {journal} {\bibinfo  {journal}
  {Phys. Rev. Lett.}\ }\textbf {\bibinfo {volume} {131}},\ \bibinfo {pages}
  {061401} (\bibinfo {year} {2023}{\natexlab{b}})},\ \Eprint
  {http://arxiv.org/abs/2305.11207} {arXiv:2305.11207} \BibitemShut {NoStop}%
\bibitem [{\citenamefont {Johns}(2023)}]{johns2023thermodynamics}%
  \BibitemOpen
  \bibfield  {author} {\bibinfo {author} {\bibfnamefont {L.}~\bibnamefont
  {Johns}},\ }\href@noop {} {\  (\bibinfo {year} {2023})},\ \Eprint
  {http://arxiv.org/abs/2306.14982} {arXiv:2306.14982} \BibitemShut {NoStop}%
\bibitem [{\citenamefont {Johns}(2024)}]{johns2024subgrid}%
  \BibitemOpen
  \bibfield  {author} {\bibinfo {author} {\bibfnamefont {L.}~\bibnamefont
  {Johns}},\ }\href@noop {} {\  (\bibinfo {year} {2024})},\ \Eprint
  {http://arxiv.org/abs/2401.15247} {arXiv:2401.15247 [astro-ph.HE]}
  \BibitemShut {NoStop}%
\bibitem [{\citenamefont {Fiorillo}\ and\ \citenamefont
  {Raffelt}(2024{\natexlab{a}})}]{fiorillo2024fast}%
  \BibitemOpen
  \bibfield  {author} {\bibinfo {author} {\bibfnamefont {D.~F.~G.}\
  \bibnamefont {Fiorillo}}\ and\ \bibinfo {author} {\bibfnamefont {G.~G.}\
  \bibnamefont {Raffelt}},\ }\href {\doibase 10.1103/PhysRevLett.133.221004}
  {\bibfield  {journal} {\bibinfo  {journal} {Phys. Rev. Lett.}\ }\textbf
  {\bibinfo {volume} {133}},\ \bibinfo {pages} {221004} (\bibinfo {year}
  {2024}{\natexlab{a}})},\ \Eprint {http://arxiv.org/abs/2403.12189}
  {arXiv:2403.12189 [hep-ph]} \BibitemShut {NoStop}%
\bibitem [{\citenamefont {Fiorillo}\ and\ \citenamefont
  {Raffelt}(2024{\natexlab{b}})}]{fiorillo2024theory}%
  \BibitemOpen
  \bibfield  {author} {\bibinfo {author} {\bibfnamefont {D.~F.~G.}\
  \bibnamefont {Fiorillo}}\ and\ \bibinfo {author} {\bibfnamefont {G.~G.}\
  \bibnamefont {Raffelt}},\ }\href {\doibase 10.1007/JHEP08(2024)225}
  {\bibfield  {journal} {\bibinfo  {journal} {JHEP}\ }\textbf {\bibinfo
  {volume} {08}},\ \bibinfo {pages} {225} (\bibinfo {year}
  {2024}{\natexlab{b}})},\ \Eprint {http://arxiv.org/abs/2406.06708}
  {arXiv:2406.06708 [hep-ph]} \BibitemShut {NoStop}%
\bibitem [{\citenamefont {George}\ \emph {et~al.}(2023)\citenamefont {George},
  \citenamefont {Lin}, \citenamefont {Wu}, \citenamefont {Liu},\ and\
  \citenamefont {Xiong}}]{george2023cosenu}%
  \BibitemOpen
  \bibfield  {author} {\bibinfo {author} {\bibfnamefont {M.}~\bibnamefont
  {George}}, \bibinfo {author} {\bibfnamefont {C.-Y.}\ \bibnamefont {Lin}},
  \bibinfo {author} {\bibfnamefont {M.-R.}\ \bibnamefont {Wu}}, \bibinfo
  {author} {\bibfnamefont {T.~G.}\ \bibnamefont {Liu}}, \ and\ \bibinfo
  {author} {\bibfnamefont {Z.}~\bibnamefont {Xiong}},\ }\href {\doibase
  10.1016/j.cpc.2022.108588} {\bibfield  {journal} {\bibinfo  {journal}
  {Comput. Phys. Commun.}\ }\textbf {\bibinfo {volume} {283}},\ \bibinfo
  {pages} {108588} (\bibinfo {year} {2023})},\ \Eprint
  {http://arxiv.org/abs/2203.12866} {arXiv:2203.12866} \BibitemShut {NoStop}%
\bibitem [{\citenamefont {{Mezzacappa}}\ and\ \citenamefont
  {{Bruenn}}(1993)}]{mezzacappa1993numerical}%
  \BibitemOpen
  \bibfield  {author} {\bibinfo {author} {\bibfnamefont {A.}~\bibnamefont
  {{Mezzacappa}}}\ and\ \bibinfo {author} {\bibfnamefont {S.~W.}\ \bibnamefont
  {{Bruenn}}},\ }\href {\doibase 10.1086/172395} {\bibfield  {journal}
  {\bibinfo  {journal} {Astrophys. J.}\ }\textbf {\bibinfo {volume} {405}},\
  \bibinfo {pages} {669} (\bibinfo {year} {1993})}\BibitemShut {NoStop}%
\bibitem [{\citenamefont {{Liebend{\"o}rfer}}\ \emph
  {et~al.}(2001)\citenamefont {{Liebend{\"o}rfer}}, \citenamefont
  {{Mezzacappa}},\ and\ \citenamefont
  {{Thielemann}}}]{liebendorfer2001conservative}%
  \BibitemOpen
  \bibfield  {author} {\bibinfo {author} {\bibfnamefont {M.}~\bibnamefont
  {{Liebend{\"o}rfer}}}, \bibinfo {author} {\bibfnamefont {A.}~\bibnamefont
  {{Mezzacappa}}}, \ and\ \bibinfo {author} {\bibfnamefont {F.-K.}\
  \bibnamefont {{Thielemann}}},\ }\href {\doibase 10.1103/PhysRevD.63.104003}
  {\bibfield  {journal} {\bibinfo  {journal} {Phys. Rev. D}\ }\textbf {\bibinfo
  {volume} {63}},\ \bibinfo {eid} {104003} (\bibinfo {year} {2001})},\ \Eprint
  {http://arxiv.org/abs/astro-ph/0012201} {arXiv:astro-ph/0012201 [astro-ph]}
  \BibitemShut {NoStop}%
\bibitem [{\citenamefont {{Liebend{\"o}rfer}}\ \emph
  {et~al.}(2004)\citenamefont {{Liebend{\"o}rfer}}, \citenamefont {{Messer}},
  \citenamefont {{Mezzacappa}}, \citenamefont {{Bruenn}}, \citenamefont
  {{Cardall}},\ and\ \citenamefont {{Thielemann}}}]{liebendorfer2004finite}%
  \BibitemOpen
  \bibfield  {author} {\bibinfo {author} {\bibfnamefont {M.}~\bibnamefont
  {{Liebend{\"o}rfer}}}, \bibinfo {author} {\bibfnamefont {O.~E.~B.}\
  \bibnamefont {{Messer}}}, \bibinfo {author} {\bibfnamefont {A.}~\bibnamefont
  {{Mezzacappa}}}, \bibinfo {author} {\bibfnamefont {S.~W.}\ \bibnamefont
  {{Bruenn}}}, \bibinfo {author} {\bibfnamefont {C.~Y.}\ \bibnamefont
  {{Cardall}}}, \ and\ \bibinfo {author} {\bibfnamefont {F.~K.}\ \bibnamefont
  {{Thielemann}}},\ }\href {\doibase 10.1086/380191} {\bibfield  {journal}
  {\bibinfo  {journal} {Astrophys. J., Suppl.}\ }\textbf {\bibinfo {volume}
  {150}},\ \bibinfo {pages} {263} (\bibinfo {year} {2004})},\ \Eprint
  {http://arxiv.org/abs/astro-ph/0207036} {arXiv:astro-ph/0207036 [astro-ph]}
  \BibitemShut {NoStop}%
\bibitem [{\citenamefont {Zaizen}\ and\ \citenamefont
  {Nagakura}(2024)}]{zaizen2023fast}%
  \BibitemOpen
  \bibfield  {author} {\bibinfo {author} {\bibfnamefont {M.}~\bibnamefont
  {Zaizen}}\ and\ \bibinfo {author} {\bibfnamefont {H.}~\bibnamefont
  {Nagakura}},\ }\href {\doibase 10.1103/PhysRevD.109.083031} {\bibfield
  {journal} {\bibinfo  {journal} {Phys. Rev. D}\ }\textbf {\bibinfo {volume}
  {109}},\ \bibinfo {pages} {083031} (\bibinfo {year} {2024})},\ \Eprint
  {http://arxiv.org/abs/2311.13842} {arXiv:2311.13842 [astro-ph.HE]}
  \BibitemShut {NoStop}%
\bibitem [{\citenamefont {Hunter}(2007)}]{matplotlib}%
  \BibitemOpen
  \bibfield  {author} {\bibinfo {author} {\bibfnamefont {J.~D.}\ \bibnamefont
  {Hunter}},\ }\href {\doibase 10.1109/MCSE.2007.55} {\bibfield  {journal}
  {\bibinfo  {journal} {Computing in Science \& Engineering}\ }\textbf
  {\bibinfo {volume} {9}},\ \bibinfo {pages} {90} (\bibinfo {year}
  {2007})}\BibitemShut {NoStop}%
\bibitem [{\citenamefont {van~der Walt}\ \emph {et~al.}(2011)\citenamefont
  {van~der Walt}, \citenamefont {Colbert},\ and\ \citenamefont
  {Varoquaux}}]{numpy}%
  \BibitemOpen
  \bibfield  {author} {\bibinfo {author} {\bibfnamefont {S.}~\bibnamefont
  {van~der Walt}}, \bibinfo {author} {\bibfnamefont {S.~C.}\ \bibnamefont
  {Colbert}}, \ and\ \bibinfo {author} {\bibfnamefont {G.}~\bibnamefont
  {Varoquaux}},\ }\href {\doibase 10.1109/MCSE.2011.37} {\bibfield  {journal}
  {\bibinfo  {journal} {Computing in Science \& Engineering}\ }\textbf
  {\bibinfo {volume} {13}},\ \bibinfo {pages} {22} (\bibinfo {year}
  {2011})}\BibitemShut {NoStop}%
\bibitem [{\citenamefont {Fischer}\ \emph {et~al.}(2010)\citenamefont
  {Fischer}, \citenamefont {Whitehouse}, \citenamefont {Mezzacappa},
  \citenamefont {Thielemann},\ and\ \citenamefont
  {Liebend{\"{o}}rfer}}]{fischer2010protoneutron}%
  \BibitemOpen
  \bibfield  {author} {\bibinfo {author} {\bibfnamefont {T.}~\bibnamefont
  {Fischer}}, \bibinfo {author} {\bibfnamefont {S.~C.}\ \bibnamefont
  {Whitehouse}}, \bibinfo {author} {\bibfnamefont {A.}~\bibnamefont
  {Mezzacappa}}, \bibinfo {author} {\bibfnamefont {F.-K.~K.}\ \bibnamefont
  {Thielemann}}, \ and\ \bibinfo {author} {\bibfnamefont {M.}~\bibnamefont
  {Liebend{\"{o}}rfer}},\ }\href {\doibase 10.1051/0004-6361/200913106}
  {\bibfield  {journal} {\bibinfo  {journal} {Astron. Astrophys.}\ }\textbf
  {\bibinfo {volume} {517}},\ \bibinfo {pages} {A80} (\bibinfo {year}
  {2010})},\ \Eprint {http://arxiv.org/abs/0908.1871} {arXiv:0908.1871}
  \BibitemShut {NoStop}%
\bibitem [{\citenamefont {{Fischer}}\ \emph {et~al.}(2020)\citenamefont
  {{Fischer}}, \citenamefont {{Guo}}, \citenamefont {{Mart{\'i}nez-Pinedo}},
  \citenamefont {{Liebend{\"o}rfer}},\ and\ \citenamefont
  {{Mezzacappa}}}]{fischer2020muonization}%
  \BibitemOpen
  \bibfield  {author} {\bibinfo {author} {\bibfnamefont {T.}~\bibnamefont
  {{Fischer}}}, \bibinfo {author} {\bibfnamefont {G.}~\bibnamefont {{Guo}}},
  \bibinfo {author} {\bibfnamefont {G.}~\bibnamefont {{Mart{\'i}nez-Pinedo}}},
  \bibinfo {author} {\bibfnamefont {M.}~\bibnamefont {{Liebend{\"o}rfer}}}, \
  and\ \bibinfo {author} {\bibfnamefont {A.}~\bibnamefont {{Mezzacappa}}},\
  }\href {\doibase 10.1103/PhysRevD.102.123001} {\bibfield  {journal} {\bibinfo
   {journal} {Phys. Rev. D}\ }\textbf {\bibinfo {volume} {102}},\ \bibinfo
  {eid} {123001} (\bibinfo {year} {2020})},\ \Eprint
  {http://arxiv.org/abs/2008.13628} {arXiv:2008.13628 [astro-ph.HE]}
  \BibitemShut {NoStop}%
\end{thebibliography}
%merlin.mbs apsrev4-1.bst 2010-07-25 4.21a (PWD, AO, DPC) hacked
%Control: key (0)
%Control: author (72) initials jnrlst
%Control: editor formatted (1) identically to author
%Control: production of article title (-1) disabled
%Control: page (0) single
%Control: year (1) truncated
%Control: production of eprint (0) enabled
%

\clearpage
\appendix
\onecolumngrid

\begin{center}
\textbf{\large Robust integration of fast flavor conversions in classical neutrino transport\\
\it{Supplemental Material}}
\end{center}

\vspace{0.1in}
\twocolumngrid

\section{Numerical setup and simulation time}\label{SM}
We use the extended version of \textsc{cose}$\nu$ adopted in Ref.~\cite{xiong2024fast} to numerically solve the neutrino quantum kinetic equation ($\nu$QKE) and the effective classical transport (ECT) described in the main text. 
For both cases, we set the inner boundary at $20$~km and the outer boundary at $80$~km. 
Neutrino distributions of different flavors at the inner boundary are given by the equilibrium Fermi-Dirac function with the corresponding chemical potentials. 
At the outer boundary, free outflow boundary condition is employed for neutrinos propagating radially outward with $v_r>0$, while we do not inject neutrinos with $v_r<0$. 

We adopt uniform grids in discretizing the radial, angular ($v_r)$, and the neutrino energy coordinates. 
For all the $\nu$QKE and ECT runs presented in the main text, the number of the angular and energy grids are taken to be $N_{v_r}=100$ and $N_E=15$. 
In the $\nu$QKE simulations, the radial grid numbers taken depend on the adopted attenuation factor for neutrino-neutrino Hamiltonian, and we use $N_r=25,000$ and $50,000$ for the ``$\nu$QKE-L'' (with attenuation factor $a=10^{-3}$) and ``$\nu$QKE-H'' (with $a=4\times 10^{-3}$) cases, respectively. 
As discussed in the main text, the needed radial grid numbers in the ECT simulations can be significantly lower and are taken to be $N_r=250$ for all cases. 
The much reduced grid number in the ECT cases also allow us to take a fixed time step size $\Delta t=32$~ns, which is much larger than $\Delta t=3.2$~ns (1.6~ns) in Model $\nu$QKE-L ($\nu$QKE-H) used in the $\nu$QKE runs. 

For all $\nu$QKE and the ECT runs, we adopt the same effective vacuum mixing parameters $\delta m^2=8\times 10^{-5}$~eV$^2$ and $\theta_V=10^{-6}$ as in most cases of \cite{xiong2024fast}. 
For the small value of the effective mixing angle, it is used to account for the matter suppression effect since we do not explicitly include the neutrino-matter forward scattering Hamiltonian in the presented simulations.
For the collision terms accounting for the neutrino absorption and emission, their scattering with nucleons and $e^\pm$, the (inverse) nucleon-nucleon Bremsstrahlung, and the $\nu\bar\nu\leftrightarrow e^+e^-$ process; see \cite{xiong2024fast} for details. 

\begin{sfigure}[t]
\begin{centering}
\includegraphics[width=\columnwidth]{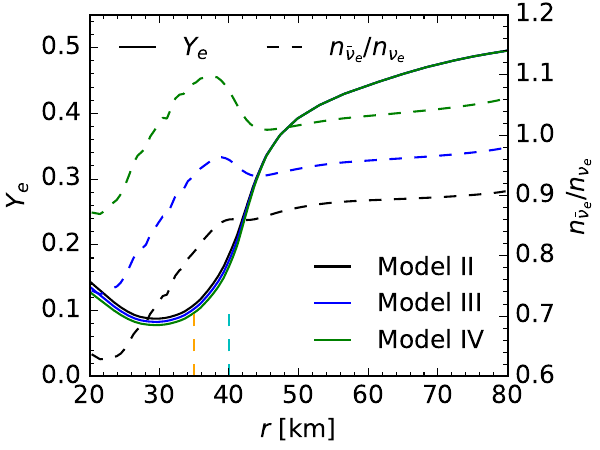}
\end{centering}
\caption{\label{fig:Ye_nratio}
The radial profiles of electron fraction $Y_e$ (dashed lines) and the number density ratio $n_{\bar\nu_e}/n_{\nu_e}$ (solid lines) in Model II (black), III (blue), and IV (green) without flavor conversions. 
The radii of the $\nu_e$ and $\bar\nu_e$ spheres are indicated by the vertical cyan and orange dashed lines, respectively.}
\end{sfigure}

The ECT simulations take essentially the same amount of time as the corresponding pure classical simulations without considering any flavor oscillations ($\sim 10$ CPU-hours on a single node machine with Intel Xeon Gold 64-core CPUs). 
However, for the $\nu$QKE simulations, even with the attenuated neutrino-neutrino forward scattering Hamiltonian, the $\nu$QKE-L and $\nu$QKE-H models require $\sim 8,000$ and $\sim 27,500$ CPU-hours on the same machine, correspondingly.

\section{Descriptions of different models}
The three supernova (SN) background models (Model II, III, and IV) used in this work are taken from Ref.~\cite{xiong2024fast}, which are based on a snapshot taken at the 252~ms post the core bounce from the spherically symmetric SN simulation of a 25~$M_\odot$ progenitor obtained with the \textsc{agile-Boltztran} code~\cite{liebendorfer2004finite,fischer2010protoneutron,fischer2020muonization} (see Ref.~\cite{xiong2024fast} for detailed description). 
Since the angular distribution crossing of the neutrino electron lepton number (ELN) that can exist in multidimensional SN simulations~\cite{azari2020fast,abbar2020fast,nagakura2019fast,glas2020fast,nagakura2021occurrence,abbar2021characteristics,harada2022prospects} are typically not present in the spherically symmetric case, 
we follow the strategy of \cite{nagakura2023basic,nagakura2023roles,xiong2024fast} by multiplying a radial dependent factor 
\begin{sequation}
    b_{Y_e}(r) = 1-\frac{b_1}{1+e^{(r/\mathrm{km})-42}},
\end{sequation}to the original electron fraction ($Y_e$) profile to generate the ELN crossings. 
We take $b_1=0.1$, $0.15$, and $0.2$ for Models II, III, and IV, respectively. 
The resulting $Y_e$ profiles and the corresponding number density ratio between $n_{\bar\nu_e}$ and $n_{\nu_e}$ without considering flavor oscillations are shown in Fig.~\ref{fig:Ye_nratio}.
Taking a larger value of $b_1$ results in more deduction of the $Y_e$ profile around the neutrinosphere, which leads to enhanced $\bar\nu_e$ production. 
As a result, $n_{\nu_e}\gtrsim n_{\bar\nu_e}$ at all radii in Models II and III, while $n_{\nu_e}\lesssim n_{\bar\nu_e}$ in Model IV. 

In Fig.~\ref{fig:eln}, we show the radial profile of the ELN angular distributions of those three models. 
Clearly, the ELN angular crossing appears at a smaller radius with increasing ratio of $n_{\bar\nu_e}/n_{\nu_e}$, accompanied with the increased range and depth of the negative part of ELN angular distribution. 
In particular, the ELN angular crossings exist inside the neutrinospheres (vertical dashed lines in Fig.~\ref{fig:Ye_nratio}) in Models III and IV, which drives the collisional feedback effect discussed in the main text and in \cite{xiong2024fast}.

\begin{sfigure}[t]
\begin{centering}
\includegraphics[width=\columnwidth]{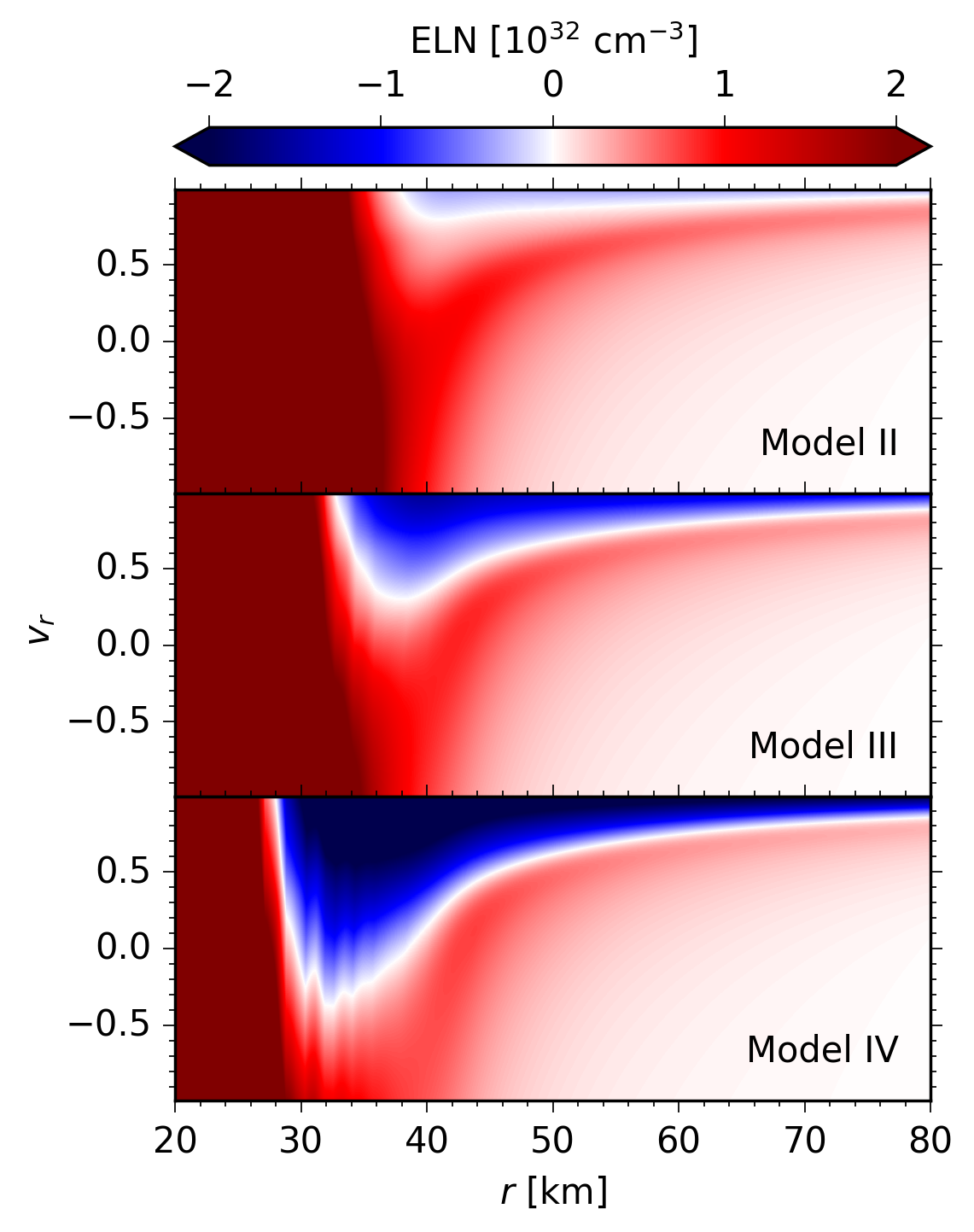}
\end{centering}
\caption{\label{fig:eln}
The ELN angular distribution as a function of radial velocity ($v_r$) and radius ($r$) in Model II--IV. 
The white band in each panel indicates where the ELN angular crossing occurs.
}
\end{sfigure}

\section{ECT results with different angular grid number and time step size}

In the main text, we have taken $N_{v_r}=100$ and $\Delta t=32$~ns for the purpose of demonstrating that the ECT scheme is able to accurately capture the effect of fast flavor conversions (FFC). 
Here, we perform two additional sets of calculations to explore the robustness of the ECT models when using reduced $N_{v_r}$ or applying the flavor redistribution prescription at time intervals larger than $\Delta t$. 
For the first set, we simply take $N_{v_r}=20$ while keeping all other settings identical as in the fiducial ECT runs. 
For the second set, since the explicit time integration scheme adopted in \textsc{cose}$\nu$~\cite{george2023cosenu,xiong2024fast} prevents us directly taking a much larger time step size, we adopt an alternative method by applying the analytical prescription to redistribute the neutrino flavor content after every $N_t=3$ or $30$ time steps to effectively explore the impact of taking a larger time step size, which may be realized by using other implicit time integration schemes. 
The case with $N_t=1$ corresponds to the fiducial ECT model presented in the main text.

Fig.~\ref{fig:supplemental_rp} compares the asymptotic radial profile of the $\nu_e$, $\bar\nu_e$, and $\nu_x$ neutrino number density ratios obtained using 
$(N_{v_r}, N_t)=(100,1)$, $(20,1)$, $(100,3)$ and $(100,30)$ 
for the Model II-ECT, III-ECT, and IV-ECT in the left, middle, and right panels, respectively. 
Compared to the fiducial cases, taking a smaller $N_{v_r}=20$ reproduces well the profiles of number density ratios in Models III and IV as well as in the decoupling region of Model II.
Minor differences ($\lesssim 5\%$) only exist at the free-streaming regime ($r\gtrsim 60$~km) outside the neutrinospheres in Model II, where neutrino angular distributions become more forward peaked, and the ELN crossings are located very close to $v_r=1$ (see Fig.~\ref{fig:eln}). 

When taking $N_t=3$, Fig.~\ref{fig:supplemental_rp} shows that the results are indistinguishable from those obtained by the fiducial ECT runs. 
Even with $N_t=30$, which effectively redistribute the neutrino flavor content only every $0.96$~$\mu$s, 
the results are nearly identical to those obtained in the fiducial runs, 
except for the region around $r\sim 34$~km in Model IV.
The underlying reason for the small difference introduced here is similar to what discussed in the main text for the ``$\nu$QKE-L'' $\nu$QKE runs. 
The too-large value of $N_t$ plays the role that effectively suppresses the FFC time scale for flavor redistribution, hence affecting the collisional feedback and causing the qualitatively similar feature as shown in Fig.~1 of the main text.

The very good agreement obtained by using a smaller value of $N_{v_r}$ comparable to the typical values used in the discrete ordinates method,
or by taking a larger $N_t$, which applies the flavor redistribution in time intervals 
on similar order as the  
time step sizes in SN simulations 
shows that the ECT scheme may be used in realistic hydrodynamical simulations without adding much extra computational burden to the neutrino transport solver.

\begin{sfigure*}[t]
\begin{centering}
\includegraphics[width=\textwidth]{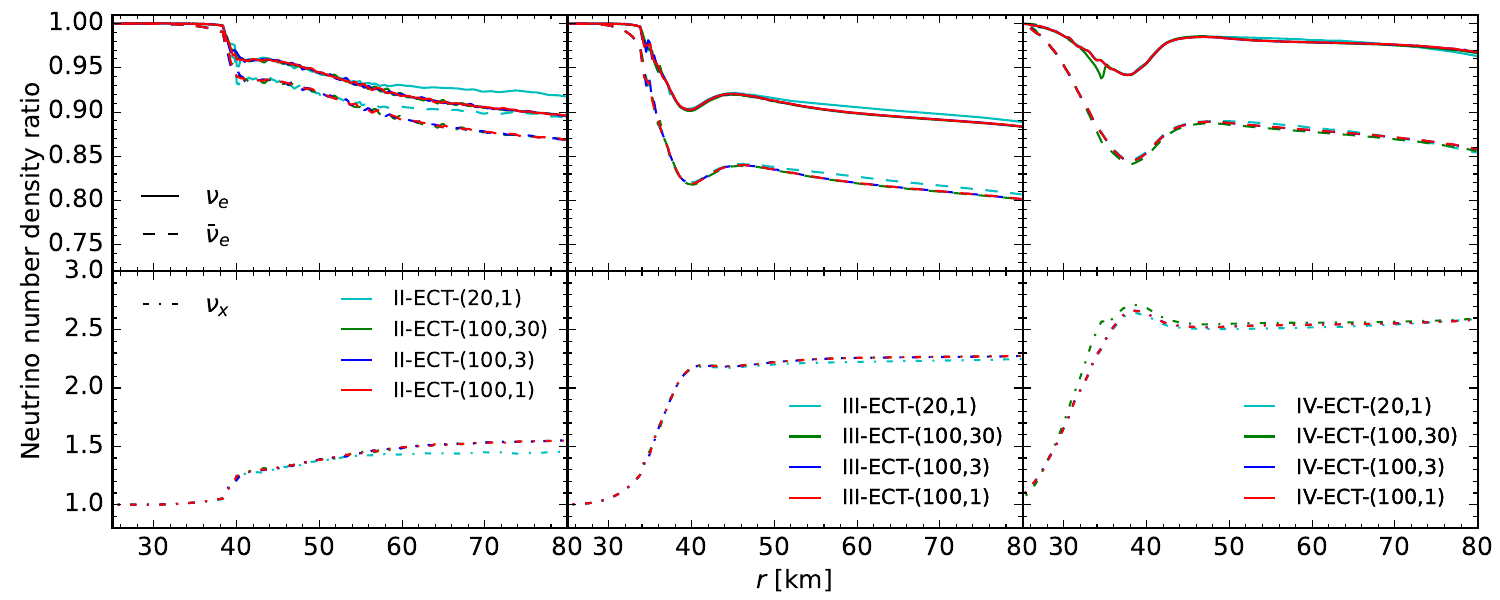}
\end{centering}
\caption{\label{fig:supplemental_rp}
Comparison of the radial profiles of neutrino number density ratios $n_\nu^{\rm FFC}/n_\nu^{\rm NFC}$ obtained in the ECT simulations taking different values of $(N_{v_r}, N_t)=(20,1)$ (cyan), $(100,30)$ (green), $(100,3)$ (blue), and $(100,1)$ (red) for Model II-ECT, III-ECT, and IV-ECT in the left, middle, and right panels, respectively.
}
\end{sfigure*}

\end{document}